\documentclass[useAMS,usenatbib,usegraphicx]{mn2e}
\usepackage{natbib}
\usepackage{times}
\usepackage{longtable}
\usepackage[T1]{fontenc}
\usepackage{aecompl}

\def\ltsima{$\; \buildrel < \over \sim \;$}
\def\simlt{\lower.5ex\hbox{\ltsima}}
\def\gtsima{$\; \buildrel > \over \sim \;$}
\def\simgt{\lower.5ex\hbox{\gtsima}}

\title[The Imprint of Reionization on the Star Formation Histories of Dwarf Galaxies]{The Imprint of Reionization on the Star Formation Histories of Dwarf Galaxies}
\author[A. Ben\'itez Llambay et al.]{A. Ben\'itez-Llambay$^{1}$\thanks{E-mail:
alejandrobll@oac.uncor.edu}, J. F. Navarro$^{2}$, M. G. Abadi$^{1}$, S. Gottl\"ober$^{3}$,
\& \newauthor G. Yepes$^{4}$, Y. Hoffman$^{5}$, M. Steinmetz $^{3}$\\
$^{1}$Observatorio Astron\'omico, Universidad Nacional de C\'ordoba, C\'ordoba, X5000BGR, Argentina\\
$^{2}$Senior CIfAR Fellow. Department of Physics \& Astronomy, University of Victoria, Victoria, BC, V8P 5C2, Canada\\
$^{3}$Leibniz Institute for Astrophysics, An der Sternwarte 16, 14482 Potsdam, Germany\\
$^{4}$Departamento de F\'isica Te\'orica, Universidad Aut\'onoma de Madrid, 28049, Madrid, Spain\\
$^{5}$Racah Institute of Physics, The Hebrew University of Jerusalem, Jerusalem, 91904, Israel}

\begin{document}

\date{}


\maketitle

\label{firstpage}

\begin{abstract}
  We explore the impact of cosmic reionization on nearby isolated dwarf galaxies using a compilation of star formation histories estimated from deep HST data and a cosmological hydrodynamical simulation of the Local Group. The nearby dwarfs show a wide diversity of star formation histories; from ancient systems that have largely completed their star formation $\sim 10$ Gyr ago to young dwarfs that have formed the majority of their stars in the past $\sim 5$ Gyr to two-component systems characterized by the overlap of comparable numbers of old and young stars. Taken as an ensemble, star formation in nearby dwarfs dips to lower-than-average rates at intermediate times ($4<t$/Gyr $<8$), a feature that we trace in the simulation to the effects of cosmic reionization.  Reionization heats the gas and drives it out of the shallow potential wells of low mass halos, affecting especially those below a sharp mass threshold that corresponds to a virial temperature of $\sim 2 \times 10^4 $ $\mathrm{K}$ at $z_{\rm reion}$.
 The loss of baryons leads to a sharp decline in the star forming activity of early-collapsing systems, which, compounded by feedback from early star formation, empties halos of gas and leaves behind systems where a single old stellar component prevails. In halos below the threshold at $z_{\rm reion}$, reionization heating leads to a delay in the onset of star formation that lasts until the halo grows massive enough to allow some of the remaining gas to cool and form stars. Young stellar components therefore dominate in dwarfs whose halos assemble late and thus form few stars before reionization. Two-component systems may be traced to late mergers of individual examples of the two aforementioned cases. The relative dearth of intermediate-age stars in nearby dwarfs might thus be the clearest signature yet identified of the imprint of cosmic reionization on the star formation history of dwarf galaxies.  \end{abstract} 
 

\section{Introduction}

The role of cosmic reionization in regulating the formation of dwarf galaxies in hierarchically-clustering scenarios such as $\Lambda$ Cold Dark Matter ($\Lambda$CDM) has long been recognized \citep{Efstathiou1992}. $\Lambda$CDM predicts the early collapse of a vast number of low-mass halos massive enough, in principle, to host luminous galaxies. These exceed by orders of magnitude the observed number of dwarfs \citep{White1978,Klypin1999,Moore1999}, hinting at a mechanism that prevents galaxies from forming altogether in the majority of these systems. Cosmic reionization, powered presumably by energetic photons that escape the formation sites of primeval galaxies and supermassive black holes, provides an obvious candidate mechanism. It quickly heats the gas to temperatures of the order of the ionization potential of atomic hydrogen (i.e., $\sim 10^4$ $\mathrm{K}$), and it can do so over large volumes of space, ``evaporating'' gas from systems whose virial\footnote{We define the virial mass, $M_{200}$, 
of a halo as that enclosed by a sphere of mean density $200$ times the critical density of the Universe, $\rho_{\rm crit}=3H^2/8\pi G$. Virial quantities are defined at that radius, and identified by a ``200'' subscript. The virial temperature is defined, for primordial mean molecular weight, by $T_{200}=36 \, (V_{200}/$km s$^{-1})^2\, \mathrm{K}$, where $V_{200}$ is the virial circular velocity. We use the parameters listed in  Sec.~\ref{SecCLUES} for all cosmology-dependent quantities.} temperatures are below that value \citep{Shapiro1994,Barkana1999} and rendering them effectively ``dark''.

In simple terms, cosmic reionization introduces two physical scales in the process of galaxy formation; a characteristic time (the redshift when reionization is effectively complete, $z_{\rm reion}$) and a characteristic mass, corresponding to that of a halo with virial temperature $\sim 10^4\, \mathrm{K}$ at $z_{\rm reion}$ ($\sim 1.2 \times 10^8 \, M_\odot$ for $z_{\rm reion}=6$). Halos that have grown substantially more massive than this characteristic mass by $z_{\rm reion}$ are affected little \citep{Thoul1996,Katz1996,Navarro1997,Okamoto2009}, whereas star formation might be heavily regulated, or even truncated, in lower mass systems \citep{Gnedin2000,Hoeft2006,Okamoto2008}. These ideas have proven essential to the success of galaxy formation models that attempt to match the faint end of the galaxy luminosity function \citep{Benson2002,Somerville2002}, or to explain the scarcity of luminous satellites in the vicinity of the Milky Way \citep{Bullock2000}.

Given the preeminent role ascribed to reionization in regulating the evolution of dwarf galaxies, it is somewhat surprising that no ``smoking gun'' evidence has yet emerged regarding its role in shaping their star formation history \citep{Grebel2004}. This is due in part to the limited guidance afforded by theoretical models, which have made disparate predictions, ranging from a complete suppression of star formation just after reionization \citep{Ricotti2005} to a slow decline in star formation rates (SFRs) as reionization-heated gas flows out of low mass halos and starves galaxies of star formation fuel \citep{Okamoto2008,Sawala2010} to late ($z\simlt 2$) outbursts in star formation activity as the intensity of the UV background decreases \citep{Babul1992,Ricotti2009}.

From an observational standpoint, deep imaging with the Hubble Space Telescope (HST) has been combined with sophisticated stellar evolution models in order to derive the star formation history (hereafter SFH, for short) of a number of nearby dwarfs \citep[see, e.g.,][and references therein]{Weisz2011}. The models constrain the age distribution and metal enrichment history of stars in a dwarf by carefully matching theoretical isochrones to the color-magnitude diagram of each galaxy. A number of algorithms have been developed for this task over the years, but a number of recent papers show that most yield comparable age and metallicity distributions when applied to data of similar quality \citep{Tolstoy2009}. The modeling of dwarf galaxy SFHs thus seems mature, although it still suffers from sizeable uncertainties when data are affected by crowding (especially in fields near the centre of a dwarf), small spatial coverage (many dwarfs are larger than the field of view of HST), or limited depth (models do best 
when the 
faint main sequence turnoff of the oldest population is reached).

We shall focus here on results obtained for $46$ dwarfs observed with HST (i.e., galaxies fainter than an absolute B-band magnitude $M_B\sim -16$, or, equivalently, with stellar masses below $10^9\, M_\odot$) located within $\sim 4.5$ Mpc from the Local Group barycenter, but {\it excluding} the satellites of bright galaxies such as the Milky Way, Andromeda, M81, M82 and NGC 2403. Satellite galaxies (defined here as those within $300$ kpc from any of those galaxies) are likely affected by tidal and halo ram pressure forces, which add extra complexity to the interpretation of their evolution. We therefore defer their analysis to a later paper and focus here on the evolution of isolated dwarfs.

The relatively large sample of systems with available  star formation histories have led to the identification of a number of robust trends. One of the most striking features is its sheer diversity: some dwarfs have been forming stars at a nearly constant rate for a Hubble time \citep[e.g., IC 1613;][]{Skillman2014}; others have mostly old stellar populations \citep[e.g., Cetus and Tucana;][]{Monelli2010a}; and some have formed most of their stars in the past few Gyrs at rates that far exceed their past average \citep[e.g., KDG2 and KDG73;][]{Weisz2011}. This diversity has been taken to suggest that reionization has not played a dominant role in the star formation history of dwarfs, since in that case one would have expected a more homogeneous population \citep{Grebel2004}.

It is also clear that essentially all isolated dwarfs, even ones where only an old stellar population is present, have formed stars for a protracted period substantially longer\footnote{We note that \citet{Brown2012} argue for a single star formation episode of short duration in some of the ultra-faint dwarfs, but these are satellites of the Milky Way and not isolated objects.}  than the short $\sim 1$ Gyr timescale elapsed before the Universe became fully reionized (recall that in $\Lambda$CDM $z=6$ corresponds to a cosmic time $t \sim 1 $ Gyr since the Big Bang).  Supporting evidence comes not only from Color-Magnitude Diagram (CMD) analyses, but also from the lack of sharp truncations in the metallicity distribution of most dwarfs, as expected if reionization had led to an abrupt cessation of star formation.

Neither of these observations, however, is necessarily incompatible with a prominent role for reionization. Indeed, the diversity of star formation histories can be reproduced in semi-analytic models \citep[see, e.g.,][]{Orban2008,Starkenburg2013}, but the models are crude, reflecting our poor understanding of the star formation process in general. In such models, erratic star formation activity typically results from sharp thresholds in the local surface gas density required for star formation. This affects low mass galaxies in particular, since it allows low density gas to accumulate in a galaxy, where it can fuel discrete star formation episodes triggered even by minor accretion events.

\citet{Ricotti2005} have also argued, on the basis of numerical simulations, that diversity in dwarf SFHs might actually be a relic feature of reionization. In their proposal,  dwarfs whose star forming history was sharply truncated early on are ``true fossils'' of reionization, but those that continued forming stars, depending on their halo mass, are either  ``polluted fossils'' or ``survivors''. Their simulations, however, evolve dwarfs only until $z=8$, and therefore  large and uncertain extrapolation is needed to contrast these predictions with observations of dwarfs in the local Universe. \citet{Shen2013} evolve a group of dwarf galaxies to $z=0$ and report ``bursty'' and diverse SFHs, but do not explore in detail the role of reionization in the origin of such diversity.

Protracted star formation and enrichment are not inconsistent with reionization either. Indeed, reionization is expected to evaporate mainly low-density gas from the halos of dwarf galaxies but to have little effect on gas that has already cooled to high-densities, where it can self-shield from the ambient UV background. This gas will continue to form stars for as long as it takes to deplete it or to expel it from the galaxy via feedback \citep{Susa2004,Hoeft2006,Okamoto2009,Sawala2012,Simpson2013}. Thus, the effect of reionization on dwarf galaxies is more likely to be a slow tapering rather than a sharp truncation of their star forming activity.

One thing that is clear from the above discussion, however, is that there is little consensus on how to identify the impact of reionization on the star formation history of dwarf galaxies. We address this issue here using cosmological simulations of the Local Group from the Constrained Local UniversE Simulations (CLUES) Collaboration \citep{Gottloeber2010} and comparing them with literature data on the SFHs of nearby dwarfs. Our analysis benefits from the excellent mass resolution of CLUES and from the specially tailored environment, which is carefully selected to match the positions, masses and velocities of the two giant spirals of the Local Group, as well as the tidal field from the nearest large-scale structures \citep{Yepes2014}. We describe the simulations briefly in Sec.~\ref{SecNumSims} and compare observations with the SFH of simulated dwarfs in Sec.~\ref{SecSFHs}. We then examine them in the context of their assembly history and of cosmic reionization in Sec.~\ref{SecAssHist} before discussing our 
results and concluding with a brief summary in Sec.~\ref{SecConc}.

\begin{figure}
 \includegraphics[width=80mm]{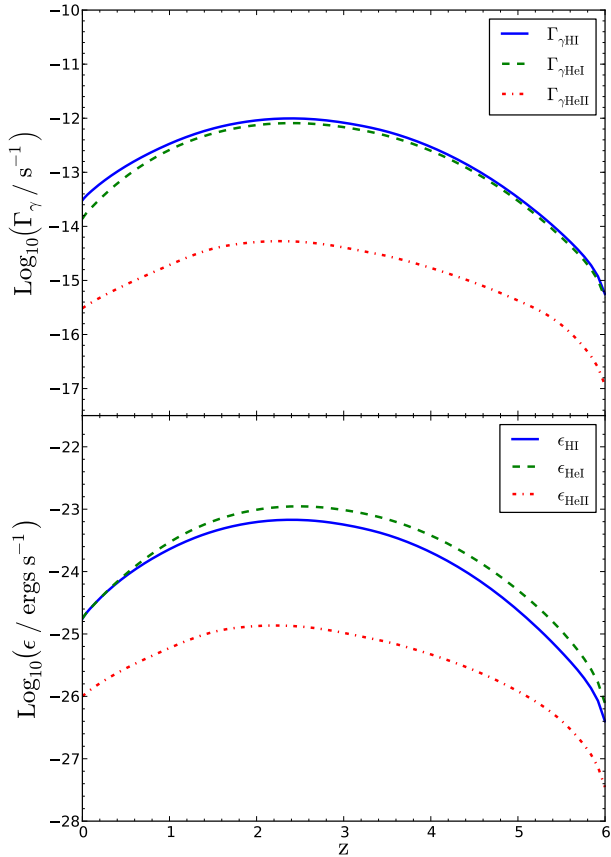}
 \caption{Photoionization (top panel) and photoheating (bottom panel) rates as a function of redshift for hydrogen and helium used in the simulation.}
\label{FigUVPar}
\end{figure}

\begin{figure}
 \includegraphics[width=84mm]{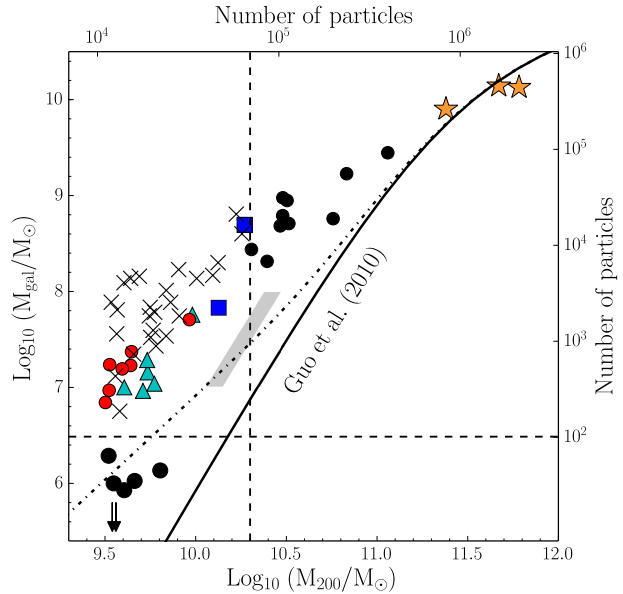}
 \caption{Stellar mass vs virial mass of galaxies in the CLUES simulation volume. We have retained for analysis all dwarfs in halos resolved with more than $\sim 10^4$ particles ($M_{200}\simgt 3 \times 10^9\, M_\odot$) but less massive than $2\times 10^{10}\, M_\odot$ (vertical dashed-line), since halos more massive than that are unlikely to be affected by reionization. A few halos in this mass range ($7$ out of $53$) have $M_{\rm gal} < 3\times 10^6 \, M_\odot$ (horizontal dashed-line), and are removed from the sample, since we cannot measure accurately a star formation history for such systems. The solid line shows the galaxy-halo mass relation from the abundance-matching analysis of \citet{Guo2010}; the dotted line indicates the correction for faint galaxies suggested by \citet{Sawala2011}; and the shaded area indicates the recent constraint derived by \citet{Brook2014}. The crosses, squares, triangles and circles identify different groups of dwarfs in our sample, according to their star formation 
history.
 See the discussion of Fig.~\ref{FigSimSFHT} for details.}
\label{FigMgalMvir}
\end{figure}

\section[]{Numerical Simulations}
\label{SecNumSims}

The simulation used in this work is part of the CLUES project. CLUES runs evolve realizations of the local Universe that mimic their large-scale mass distribution and the tidal fields operating on them.  These simulations have been presented in \citet{Gottloeber2010} and different aspects of them have been reported in recent work \citep[eg.,][]{Libeskind2010, DiCintio2012, Knebe2011,Benitez-Llambay2013, Yepes2014}. We include here a brief description for completeness; full details may be consulted in the above references or in the webpage\footnote{{\tt www.clues-project.org}} of the project.

\begin{figure*}
\begin{center}
  \includegraphics[width=100mm]{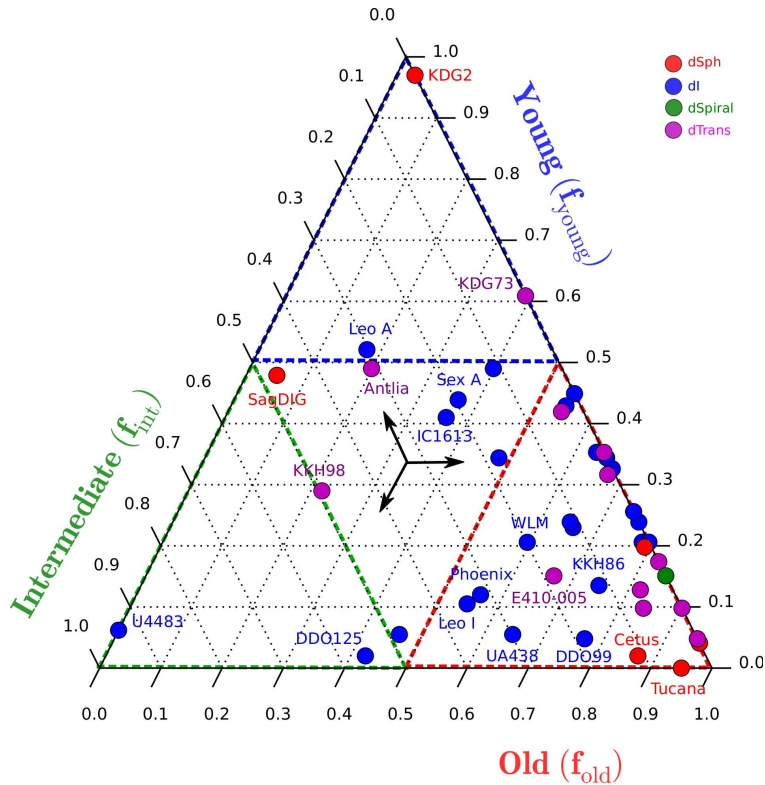}
\end{center}
 \caption{Ternary diagram summarizing the star formation histories of our sample of nearby isolated dwarf galaxies; see Table~\ref{TabObsDwarfs} for details. The three coordinates in this plot indicate the average star formation rate in the earliest $4$ Gyr (``old''), at times between $4$ and $8$ Gyr (``intermediate'') and at $t>8$ Gyr (``young''). These rates are normalized to the average star formation rate of each dwarf, $\bar f= M_{\rm gal}/t_0$ and they add up to unity. The central point in this diagram correspond to a galaxy with constant star formation rate, for which $f_{\rm old}=f_{\rm int}=f_{\rm young}=1/3$. The arrows emanating from that point show how to read coordinates in each of the three axes. Galaxies are colored by their morphological types \citep{Weisz2011}, as labeled. The colored triangles indicate the loci in this diagram of systems where one of the three stellar components dominate. For example, the red (bottom-right) triangle highlights systems where old stars formed at an average 
rate  $50\%$ higher than the past average, i.e., $f_{\rm old}>0.5$. A similar criterion identifies the other triangular regions: blue (top) corresponds to $f_{\rm young}>0.5$; and green(bottom-left) to $f_{\rm int}>0.5$. }
  \label{FigObsSFHT}
\end{figure*}
\subsection{The Code}
\label{SecCode}

The coupled evolution of baryons and dark matter is followed using the {\tt Gadget-2} code \citep{Springel2005}.  Besides gravity, the baryonic component is affected by the following processes: gas pressure and hydrodynamical shocks; radiative cooling and heating from an evolving, uniform UV background; star formation and supernova feedback; galactic winds; and metal enrichment.  

The numerical hydrodynamic treatment is based on the Smooth Particle Hydrodynamics (SPH) technique as described by \cite{Springel2003}, where details about the star formation and feedback algorithm may also be found. Part of the stellar feedback energy is invested into driving isotropic ``winds'', modeled by gas particles ejected at $\sim 340$ km/s in random directions and decoupled from further hydrodynamic interactions until its density has dropped to $10\%$ of the gas density star formation threshold, which is set at $n_H= 0.12$ cm$^{-3}$. We list the values of the parameters associated with the star formation and feedback algorithms in Table~\ref{TabSFPar}, following the same notation of \citet{Springel2003}.

\begin{table}
\caption{Adopted values of the parameters of the \citet{Springel2003} multiphase model for star formation and feedback.}
\label{TabSFPar}
\begin{center}
\begin{tabular}{@{}cccccc}
\hline
$\beta$ & $\mathrm{T_{SN}}$ & $\mathrm{\rho_{th}}$ & $\mathrm{t_*}$ & $\mathrm{T_{c}}$ & $\mathrm{A_0}$  \\
\hline
0.1 & $\mathrm{8000 \ K}$ & $\mathrm{0.12 \ cm^{-3}}$ & $\mathrm{3 \ Gyr}$ & $\mathrm{2000 \ K}$ & 1000 \\
\end{tabular}
\end{center}
\end{table}

Cosmic reionization is modeled using a uniform, time-dependent isotropic UV background radiation field switched on at redshift $z_{\rm reion}=6$.  We adopt the \citet{Haardt1996} spectrum assuming an optically thin gas in ionization equilibrium with primordial abundances. The relevant parameters for modeling its thermal evolution are the photoionization and photoheating rates, given by:

\begin{eqnarray}
 \Gamma_{\gamma i} &=& \displaystyle\int_{\nu_i}^{\infty} \displaystyle\frac{4 \pi J(\nu)}{h \nu} \sigma_{i}(\nu) d\nu \\
 \epsilon_{i} &=& \displaystyle\int_{\nu_i}^{\infty} \displaystyle\frac{4 \pi J(\nu)}{h \nu} \sigma_{i}(\nu) (h\nu-h\nu_i)d\nu, \\ \nonumber
\end{eqnarray}

\noindent where $J_{\nu}$ is the intensity of the UV background at frequency $\nu$; $\nu_{i}$ and $\sigma_i(\nu)$ are the threshold frequency and cross section for photoionization of species $i$. (Note that in this section $h$ refers to Planck's constant.) We update the relative abundance of species and the thermal history of the gas using the resulting photoionization and photoheating rates for hydrogen and helium shown in Fig.~\ref{FigUVPar}.   

We neglect the contribution of metals to the cooling rates, as well as cooling from excitation of molecular hydrogen. The inclusion of these effects would only enhance the amount of gas available for star formation. As we note below, the CLUES simulation we analyze here likely allows too many baryons to cool and turn into stars in dwarf halos, at least compared with expectations from theoretical models tuned to reproduce the faint end of the galaxy stellar mass function. Because of this shortcoming, our results are best regarded as qualitative assessments of the relative importance of various effects, rather than as quantitative predictions of the actual properties of dwarf galaxies.

\subsection{The CLUES Local Group Simulation}
\label{SecCLUES}

Our study uses the Local Group CLUES simulation run assuming cosmological parameters compatible with the WMAP 3-year measurements \citep{Spergel2007}: $\Omega_{\rm M} = 0.24$; $\Omega_{\rm b} = 0.042$; $\Omega_{\Lambda} = 0.76$; Hubble parameter $h=0.73$; $\sigma_8 = 0.75$; and $n=0.95$. The universal baryon fraction is thus $f_{\rm  bar}=\Omega_b/\Omega_{\rm M}=0.1734$. The simulation evolves a cosmological box of side length $L_{\rm box} = 87.7 \,$ Mpc from redshift $z=100$ to $0$ and includes a high-resolution ``Local Group'' region which, at $z=0$, is roughly contained within a sphere of radius $2.7 \,$ Mpc. This region was selected to contain two massive halos with relative positions, masses, and velocities consistent with those estimated for the Milky Way and Andromeda galaxies.

The Local Group region is simulated with $\sim 5.29 \times 10^7$ dark matter particles of mass $m_{\rm drk} = 2.87 \times 10^{5} \, M_{\odot}$ and the same number of gas particles, each of mass $m_{\rm gas} = 6.06 \times 10^{4} \, M_{\odot}$ respectively. The Plummer-equivalent gravitational softening length is $\epsilon_{\rm g} \sim 137 \,$ pc, and is kept fixed in comoving coordinates. Stars are modeled as collisionless particles of mass $m_{\rm str}= 3.03 \times 10^4 \, M_{\odot}$ spawned by gas particles when they are eligible to form stars. Outside the high-resolution region, the simulation includes only low-resolution dark matter particles with particle masses that increase with increasing distance to the Local Group barycenter.

\begin{figure}
\begin{center}
 \includegraphics[width=80mm]{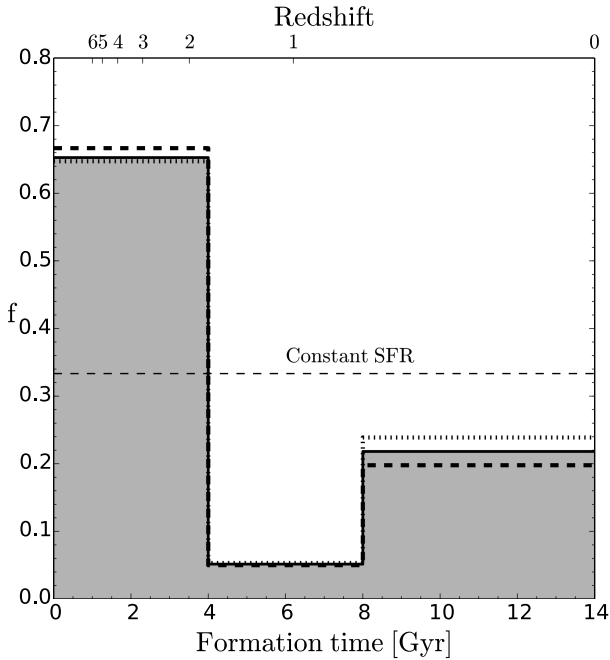}
\end{center}
 \caption{Ensemble star formation history for all observed dwarfs in our sample. The solid histogram shows the median $f_{\rm old}$, $f_{\rm int}$, and $f_{\rm young}$ coordinates of the systems shown in Fig.~\ref{FigObsSFHT}. Dotted and dashed lines are used to show the SFH of the sample, after splitting it in two, according to stellar mass. Note that, as a group, dwarfs show an intermediate age gap in their stellar populations. }
 \label{FigSFHObsDwarfs}
\end{figure}

\subsection{Simulated dwarf sample}
\label{SecSimDwarfs}

Our analysis is based on a sample of dwarf galaxies identified in a $2$ Mpc-radius sphere centered at the barycenter of the two most massive galaxies at $z=0$. These two galaxies are, by design, analogous to the Milky Way and M31: they each have a stellar mass of a few times $10^{10} \, M_\odot$ and are separated by $\sim 1.2$ $\mathrm{Mpc}$. Their relative radial velocity is $\sim -200$ km/s. We have searched for {\it isolated} dwarfs in this region using the group finder {\small SUBFIND} \citep{Springel2001}, applied to a list of friends-of-friends (FoF) halos constructed with a linking-length of $0.2$ times the mean interparticle separation. 

We retain as candidate dwarfs only those galaxies ``central'' to each FoF; i.e., we do not consider ``satellite'' galaxies. We have also checked that none of the central galaxies has been, in the past, satellite of a more massive system; in other words, we do not consider dwarfs that might have been ``ejected'' from massive halos when the groups to which they belong are disrupted tidally \citep{Balogh2000,Gill2004,Sales2009,Ludlow2010}. Our sample of simulated galaxies therefore contains, by construction, only dwarf systems that have evolved in isolation in the periphery of the massive galaxies of the Local Group.

We extend our analysis to systems with masses as low as allowed by numerical resolution. The mass per particle of our simulation is, effectively, $m_{\rm gas}+m_{\rm drk}\approx 3 \times 10^5 \, M_\odot$, which implies that a $M_{\rm lower}=3\times 10^9 \, M_\odot$ halo is realized with $\sim 10^4$ particles. We therefore choose $M_{\rm lower}$ as the minimum virial mass of the halos hosting dwarfs included in our sample.  After verifying that they are not ``contaminated'' by any of the low-resolution massive particles that provide the large-scale tidal field we measure the stellar mass, $M_{\rm gal}$, of each dwarf within a ``galactic radius'' defined as $r_{\rm gal}=0.15\, r_{200}$. We show $M_{\rm gal}$ versus the virial mass of the surrounding halos at $z=0$ in Fig.~\ref{FigMgalMvir}. 

Since our main goal is to study the impact of reionization on isolated dwarfs, we also restrict our analysis to systems with virial mass below $M_{\rm upper}=2\times 10^{10}\, M_\odot$. The progenitors of all systems more massive than $M_{\rm upper}$ had virial temperatures well above $\sim 3 \times 10^4 \, \mathrm{K}$ (or virial masses above $2 \times 10^9\, M_\odot$) at our adopted reionization redshift, $z_{\rm reion}=6$.  We do not expect reionization to play a dominant role in such systems, and therefore we do not consider them further here.

As Fig.~\ref{FigMgalMvir} makes clear, the stellar mass of simulated galaxies correlates strongly with the virial mass of their host halo, down to the smallest systems we consider in our analysis. The scatter increases steadily with decreasing halo mass: galaxy masses at $M_{\rm lower}$ vary from a maximum of $\sim 2\times 10^8 \, M_\odot$ to {\it no stars} in the case of a few halos (shown by downward-pointing arrows in Fig.~\ref{FigMgalMvir}).

The solid line in Fig.~\ref{FigMgalMvir} indicates the halo-galaxy mass relation expected from the abundance-matching model of \citet{Guo2010}. Strictly speaking, this model has only been validated observationally for $M_{\rm gal}\simgt 2 \times 10^8 \, M_\odot$ and has been extrapolated as a power law to fainter galaxies.  \citet{Sawala2011} have argued for a shallower $M_{\rm gal}$-$M_{200}$ relation at the faint end, as shown by the dotted line. However, \citet{Brook2014} have recently derived the same slope as \cite{Guo2010} but a slightly different normalization, as shown by the shaded area in Fig.~\ref{FigMgalMvir}. The simulated dwarfs are clearly above all of these models, highlighting the fact that the CLUES star formation/feedback algorithm likely allows too many stars to form in low-mass halos to be consistent with observations. Given this shortcoming, we warn that our simulations are perhaps best suited to explore qualitative trends rather than to make quantitative comparisons with individual 
galaxies.

\begin{figure}
\begin{center}
  \includegraphics[width=70mm]{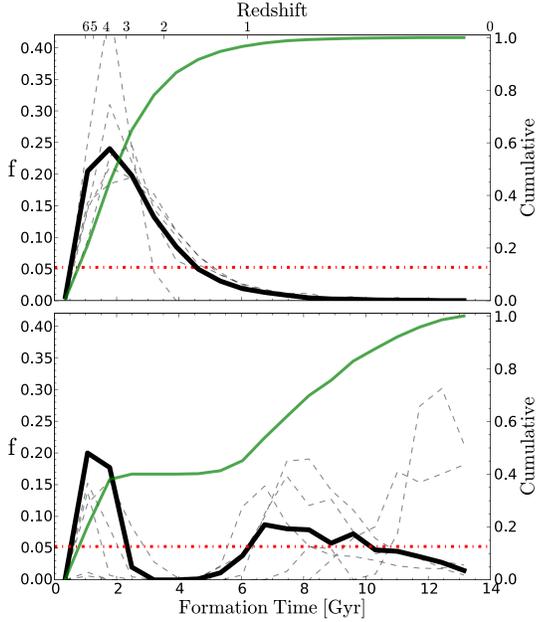}
\end{center}
 \caption{Star formation histories of simulated dwarfs in our sample, split in two groups, according to the fraction of stars formed after $t=6$ Gyr. The thick black curve in the top panel shows the average star formation rate of dwarfs that have formed less than $20 \%$ of their stars since $t=6$ Gyr; the bottom panel corresponds to the remainder of the sample. The green curve in each panel shows the average cumulative fraction of stars formed, for each group, respectively (scale on right). Thin lines in each panel show a few curves for individual systems, for illustration. Note that galaxies that form fewer than $40\%$ of their stars in the earliest $6$ Gyr show a well pronounced ``gap'' in their star formation history in the $3<t/$Gyr$<6$ regime, reminiscent of the gap in the SFH of observed dwarfs noted in Fig.~\ref{FigSFHObsDwarfs}.}
  \label{FigAvSFHSims}
\end{figure}

\section{Star formation histories}
\label{SecSFHs}

\subsection{Nearby dwarfs}
\label{SecObsDwarfs}

We have compiled a sample of nearby dwarfs with published star formation histories, mainly from the LCID\footnote{\tt http://www.iac.es/proyecto/LCID/} and the ANGST\footnote{\tt http://archive.stsci.edu/prepds/angst/} projects. Table~\ref{TabObsDwarfs} summarizes the main parameters of the dwarfs considered in our analysis, and includes all $-16 \simlt M_B \simlt -7$ ($10^{5}<M_{\rm gal}/M_\odot < 10^{9}$) dwarfs with published SFHs out to $\sim 4.5$ Mpc from the Milky Way, excluding satellites of the Milky Way, M31, M81, M82 and NGC 2403 (i.e., excluding galaxies closer than $300$ kpc from any one of those galaxies). Table~\ref{TabObsDwarfs} also lists their position in Galactic coordinates, their heliocentric radial velocity, and an estimate of the galaxy's stellar mass as listed in the quoted references.

The main data we focus on is the star formation history of a dwarf, as inferred from the age distribution of stars present at $z=0$ and taken directly from the references listed in Table~\ref{TabObsDwarfs}. We characterize these SFHs by computing, for each dwarf, the fraction of stars formed in three different intervals of cosmic time: $f_{\rm old}$ refers to ``old'' stars ($t_{\rm form}<4$ Gyr), $f_{\rm int}$ to ``intermediate-age'' stars ($4<t_{\rm form}/$Gyr$<8$), and $f_{\rm young}$ to ``young'' stars ($t_{\rm form}>8$ Gyr). 
We express these fractions as star formation rates (SFR) normalized to the past average, ${\bar f}=M_{\rm gal}/t_0$, where $t_0=13.7$ Gyr is the age of the Universe. In other words, 
\begin{equation}
f_j={1\over X}{M_ j/\Delta t_j \over {\bar f}}, 
\end{equation}
where the subscript $j$ stands for either the ``old'', ``intermediate'', or ``young'' component, and
\begin{equation}
X={1 \over {\bar f}}\sum_j M_j/\Delta t_j
\end{equation}
is a normalizing coefficient that ensures that $f_{\rm old}+f_{\rm int}+f_{\rm young}=1$.

With this definition, a dwarf that has formed stars at a constant rate has $f_{\rm young}=f_{\rm int}=f_{\rm old}=1/3$.  We show this graphically in Fig.~\ref{FigObsSFHT}, where we use a ternary diagram to show the SFH of all dwarfs in our sample. In this triangle a galaxy with constant SFR sits in the middle, and the ($f_{\rm old}$,$f_{\rm int}$,$f_{\rm young}$) coordinates are read by projecting any given point onto the three axes along lines parallel to the sides of the triangle. The arrows at the center of the plot show an example of how to read in Fig.~\ref{FigObsSFHT} the coordinates of the constant SFR point, which, by definition, are ($1/3$, $1/3$, $1/3$).

The colored areas in this plot indicate regions where one of the three populations dominates. Galaxies in the red (bottom right) triangle are dwarfs where an old stellar population prevails ($f_{\rm old}>1/2$): these galaxies formed stars in the oldest time bin at rates $\sim 50\%$ higher than expected at constant rate. An analogous definition applies to the blue (young) and green (intermediate) triangles in Fig.~\ref{FigObsSFHT}. The central triangle, on the other hand, is populated by systems where no one single population prevails.

Fig.~\ref{FigObsSFHT} summarizes, in quantitative form, the diverse star formation histories of nearby dwarfs. A few general trends are readily noticeable: (i) there is a relatively poor correlation between morphological type, denoted by the color of each point (as labeled), and this particular rendering of the star formation history of a dwarf; (ii) most dwarfs have prominent ``old'' stellar components ($71 \%$ have $f_{\rm old}>0.5$); (iii) few have dominant young or intermediate populations (i.e., only $7 \%$ have $f_{\rm young}>0.5$ and $4 \%$ have $f_{\rm int}>0.5$, respectively); and (iv) star formation during the ``intermediate'' time interval seems to have been particularly disfavoured ($\sim 76 \%$ have $f_{\rm int}<0.2$). 

The latter point is particularly clear, given the large number of systems that hug the $f_{\rm int}=0$ axis. This is interesting, since it suggests the presence of a mechanism that is able to inhibit star formation in isolated dwarfs during the period $4<t/$Gyr$<8$ (corresponding to $1.7\simgt z\simgt 0.6$). This overall hiatus in star formation activity is clearly seen in Fig.~\ref{FigSFHObsDwarfs}, where we show the median $f_{\rm young}$, $f_{\rm int}$, and $f_{\rm old}$ of all dwarfs listed in Table~\ref{TabObsDwarfs}. This figure also shows that the ``gap'' in star forming activity is not sensitive to galaxy mass: the dotted and dashed histograms in Fig.~\ref{FigSFHObsDwarfs} indicate the median star formation history of observed dwarfs, split in two bins of galaxy mass; i.e., those more and less massive than $M_{\rm gal}\sim 2\times 10^{7} \, M_\odot$, respectively. The same gap is apparent in both groups.

\subsection{Simulated dwarfs}
\label{SecSFHSims}

Guided by the preceding discussion we examine the star formation histories of the simulated dwarfs to look for clues to the origin of the decline in star formation activity at intermediate times. The thick lines in Fig.~\ref{FigAvSFHSims} indicate the average SFH of simulated dwarfs, split in two groups, according to the relative importance of star formation before and after $t=6$ Gyr (corresponding to $z \sim 1$). The top panel corresponds to systems that have formed more than $80 \%$ of their stars before that time, the bottom panel refers to the remainder of the sample. 

Interestingly, the median SFH of simulated dwarfs in the latter group (where star formation has continued until recently) show a pronounced gap between $t=3$ and $6$ Gyr, reminiscent of that seen in nearby dwarfs. Although the gap appears about $\sim 1$ Gyr earlier in the simulations than in the observations, we believe that the similarity between the bottom panel of Fig.~\ref{FigAvSFHSims} and that of Fig.~\ref{FigSFHObsDwarfs} warrants further scrutiny, especially since a $1$ Gyr shift is not large considering the typical age uncertainty of population synthesis models.

\begin{figure}
 \includegraphics[width=80mm]{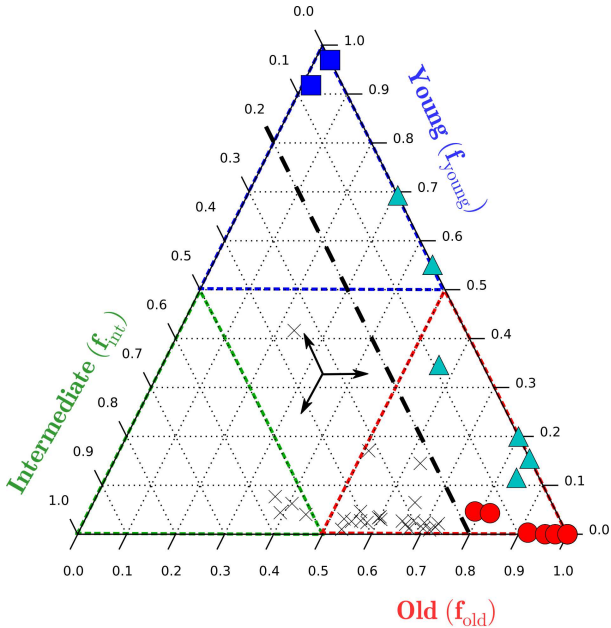}
 \caption{Same as Fig.~\ref{FigObsSFHT} but for the simulated dwarf sample. The young, intermediate, and old stellar populations here refer to formation times bracketed by $t=3$ and $6$ Gyr. Systems to the right of the black thick dashed line are those with a gap in intermediate-age stars, i.e., $f_{\rm int}<0.2$. Among these, we identify those where a single old stellar population is present (red circles, $f_{\rm young}<0.1$); where a single young stellar population dominates (blue squares, $f_{\rm old}<0.1$); and those with two separate populations of distinct age (cyan triangles).}
 \label{FigSimSFHT}
\end{figure}

We summarize in Fig.~\ref{FigSimSFHT} the star formation histories of the simulated dwarfs in our sample using a ternary diagram analogous to that shown in Fig.~\ref{FigObsSFHT}, but using the period $3<t/$Gyr$<6$ (corresponding to $2.3 \simgt z \simgt 1.0$) to define the ``intermediate'' stellar population, and adjusting accordingly the definitions of the ``old'' ($t_{\rm form}<3$ Gyr) and ``young'' ($t_{\rm form}>6$ Gyr) populations.  Fig.~\ref{FigSimSFHT} shows that simulated dwarfs exhibit a wide diversity of SFHs, akin to that of nearby dwarfs.

The prevalence of ``old'' stellar populations, however, is more pronounced in the simulation, likely as a result of the relatively inefficient feedback prescription adopted, which, as discussed in Sec.~\ref{SecSimDwarfs}, is unable to regulate effectively the total mass in stars formed in low mass halos. This is especially true at high redshift, when dwarfs assemble much of their mass and the supply of gas for star formation is plentiful. 

Furthermore, many of these galaxies have a substantial fraction of their baryons stripped by ram pressure as they move through the cosmic web \citep{Benitez-Llambay2013}. This ``cosmic web stripping'' operates most efficiently at $z\sim 2$ in the simulation we analyze here, and deprives many of these galaxies from the star formation fuel needed to continue forming stars at late times. Early star formation thus dominates over intermediate and recent in most simulated dwarfs.

\begin{figure}
\begin{center}
 \includegraphics[width=80mm]{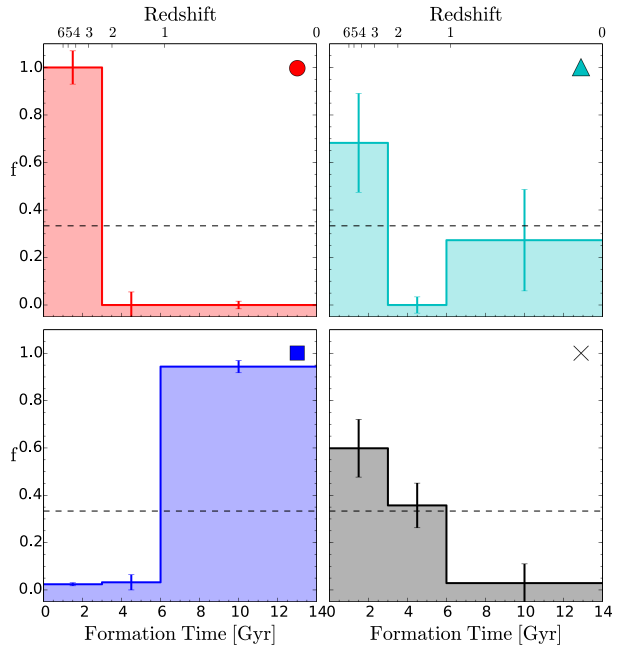}
\end{center}
 \caption{As Fig.~\ref{FigSFHObsDwarfs}, but for each of the four groups identified in Fig.~\ref{FigSimSFHT}. The three colored histograms correspond to systems with a gap in their intermediate-age stellar populations (i.e., $f_{\rm int}<0.2$). The grey-shaded histogram shows the average star formation history of the rest. The symbols in each panel identifies each group that was averaged, as in  Fig.~\ref{FigSFHObsDwarfs}. Error-bars indicate the rms dispersion per bin. See text for further details.}
 \label{FigSFHSimDwarfs}
\end{figure}

This caveat notwithstanding, Fig.~\ref{FigSimSFHT} also shows that a number of simulated dwarfs have star formation histories that resemble closely that shown in Fig.~\ref{FigSFHObsDwarfs}. In particular, some show a well defined hiatus in their star formation activity at intermediate times: only $7 \%$ of them have $f_{\rm int}>0.5$ and roughly $46\%$ of all systems have $f_{\rm int}<0.2$. The median SFH of the latter group (highlighted in color in Fig.~\ref{FigSimSFHT}) thus resembles that of observed dwarfs and their evolution might therefore help to elucidate the origin of the decline in star formation at intermediate times discussed in the previous subsection.  

We are particularly interested in systems where $f_{\rm int}$ is low but not as a result of either an overly dominant old component ($f_{\rm young} <0.1$; shown in red), or young component ($f_{\rm old}<0.1$; shown in blue). These galaxies (highlighted in cyan in Fig.~\ref{FigSimSFHT}) are those that exhibit a well-defined intermediate-age gap, and where two stellar populations of different age co-exist. The average star formation history of each of these groups is shown in Fig.~\ref{FigSFHSimDwarfs}. We use this grouping to explore below clues to the origin of these features in the star formation histories of simulated dwarfs.

\begin{figure*}
\begin{center}
 \includegraphics[width=164mm]{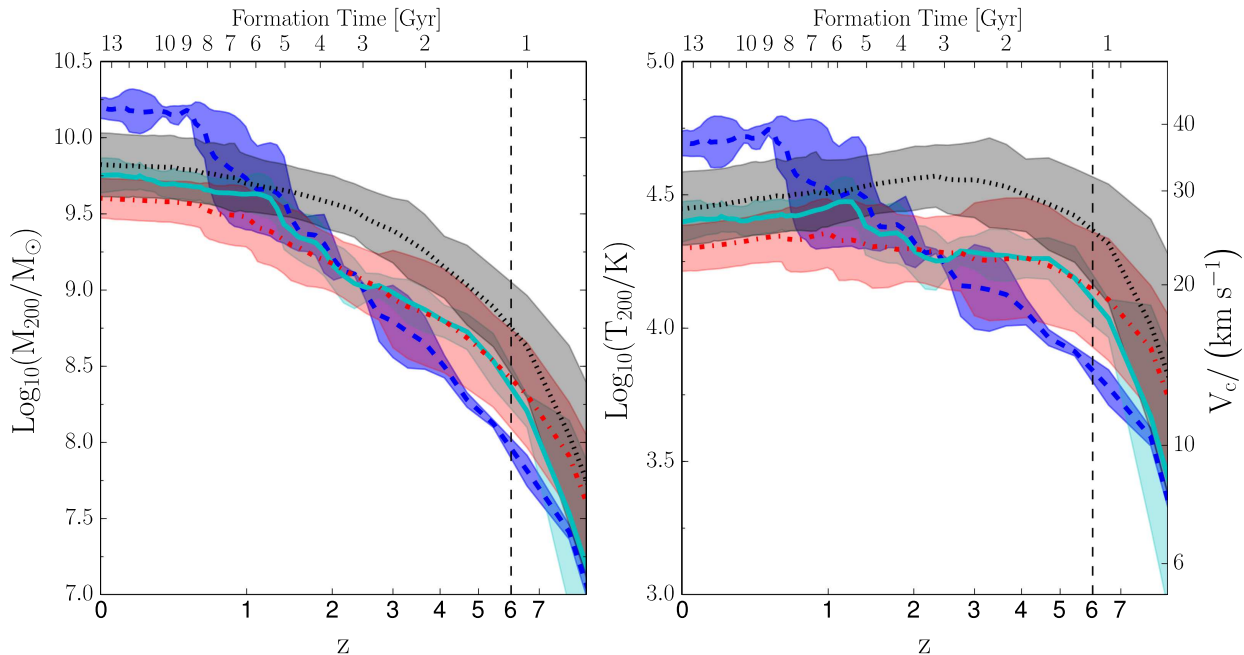}
\end{center}
 \caption{Evolution of the virial mass, $M_{200}$, (left panel) and of the virial temperature, $T_{200}$, (right panel) of the main progenitors of the simulated dwarfs, separated into four groups according to their SFH, as indicated in Fig.~\ref{FigSimSFHT}. Different lines indicate the average for each group, shaded areas the rms deviation. Note that membership to each group may be discerned from the mass (or virial temperature) of a dwarf's progenitor at $z_{\rm reion}$, but correlates poorly with virial mass at $z=0$. Those most massive at $z_{\rm reion}$ (grey dotted line) are able to initiate star formation before reionization and to sustain it for several Gyr. Less massive progenitors at $z_{\rm reion}$ are more susceptible to cosmic reionization, which induces a pronounced decline in star forming activity at intermediate times ($f_{\rm int}<0.2$). Among the latter, the least massive form stars predominantly late (blue dashed line), while the most massive experience a single star formation episode at 
very early times (red dash-dotted line). Systems with two components of distinct age (cyan solid line) are made out of a relatively recent merger between two prototypes of each of the two aforementioned groups. }
  \label{FigMTEvol}
\end{figure*}

\section{Reionization and Assembly Histories}
\label{SecAssHist}

The results of the preceding section show that, despite the fact that the simulated dwarfs in our sample have been chosen to reside in halos within a narrow range in mass, their star formation histories can still vary widely from system to system. What causes this diversity? As we show below, its origins may be traced to the mass accretion history of the dwarf and to the effects of cosmic reionization. 

We illustrate this in Fig.~\ref{FigMTEvol}, where we plot the redshift evolution of the virial mass (left panel) and of the virial temperature (right panel) of the dwarfs' main progenitors, grouped together as in Fig.~\ref{FigSimSFHT}. For each group, the thick line and the shaded area indicate the mean and the rms deviation, respectively.  Fig.~\ref{FigMTEvol} shows clearly that the star formation history of a dwarf is intimately linked to its progenitor mass at the time of reionization.

The most massive progenitors at $z_{\rm reion}$ (shown in grey) define the group of dwarfs with protracted star formation and where an intermediate-age gap is not present (i.e., those with $f_{\rm int}>0.2$; see the grey crosses in Fig.~\ref{FigSimSFHT} and the bottom-right panel in Fig.~\ref{FigSFHSimDwarfs}). These halos grow from $\sim 5\times 10^8 \, M_\odot$ at the time of reionization to $\sim 5\times 10^9 \, M_\odot$ at $z=0$. The virial temperature evolves much less in the same period, peaking at $\sim 3\times 10^4$ $\mathrm{K}$ just before $z_{\rm reion}$ and remaining roughly constant afterwards.

\begin{figure}
\begin{center}
\includegraphics[width=60mm]{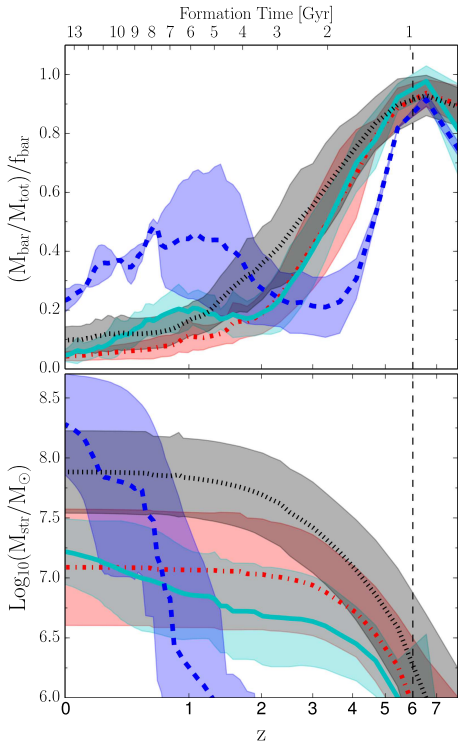}
\end{center}
 \caption{Evolution of the baryon fraction within the virial radius (upper panel) and of the stellar mass (bottom panel) for each of the four groups identified in Fig.~\ref{FigSimSFHT} according to their star formation histories. All of these galaxies lose most of their baryons as a result of cosmic reionization, aided by stellar feedback. Membership in each of the groups reflects the mass of the progenitor at $z_{\rm reion}$ (marked by the vertical dashed line), which determined how quickly baryons are evacuated from the halo of each dwarf. Those in blue correspond to the least massive progenitors at $z_{\rm reion}$. These systems are not massive enough to initiate star formation before reionization, and are therefore most severely affected when the gas is suddenly heated at $z_{\rm reion}$. This heating drives gas out of the halo and induces a delay in the onset of star formation in these systems. Those in red are massive enough to start forming stars before reionization but not enough to retain their 
baryons once reionization heating, plus the feedback of stars already formed, turn on. Systems in grey are massive enough to cool and condense enough gas before reionization to sustain star formation for a protracted period. See text for more details.}
  \label{FigFbEvol}
\end{figure}

The early collapse ensures that a substantial amount of gas in these halos is able to cool and start forming stars before $z_{\rm reion}$. When the Universe reionizes, it heats mainly the low density gas in these halos and, aided by feedback-driven winds from already-formed stars, drives it out of the virial radius, leading to a steep decline in the baryon fraction within the virial radius (see top panel of Fig.~\ref{FigFbEvol}). These processes, however, affect mostly the tenuous gas component in the halo of a dwarf, and much less the gas that has already cooled and condensed into galaxies by $z_{\rm reion}$. That gas continues to fuel star formation until it is depleted, leading to the slowly declining SFHs shown in the bottom-right panel of Fig.~\ref{FigSFHSimDwarfs}. These systems, some of which are also affected by cosmic web stripping \citep{Benitez-Llambay2013}, are today characterized by the presence of old and intermediate stellar populations of comparable importance, and a dearth of young stars.

Fig.~\ref{FigMTEvol} also highlights dwarfs with little intermediate star formation ($f_{\rm int}<0.2$; see colored symbols). These systems, most of which were at the periphery of the Local Group at $z\sim 2$ and thus less affected by cosmic web stripping, have significantly lower masses at $z_{\rm reion}$\footnote{These masses are still all above the minimum halo mass of $3 \times 10^7 \, M_{\odot}$ that~\citet{Hoeft2006} find is needed at $z_{\rm reion}$ for halos to retain at least half of their baryons (see their eq. 6)}. Their virial temperatures, in particular, fall below $2\times 10^4$ $\mathrm{K}$, making these systems much more susceptible to the heating effects of reionization. The overall result, however, depends not only $T_{\rm vir}(z_{\rm reion})$, but is also sensitive to the halo mass at late times as well as on the mass of stars formed before reionization.

Take, for example, dwarfs with a single, very old stellar population; i.e., those with $f_{\rm int}<0.2$ and $f_{\rm young} <0.1$, shown in red. Their evolution mimics that of the most massive progenitor population (shown in grey) except for their correspondingly lower masses and shallower potential wells. Together, these imply that less gas condenses before reionization, leading to fewer stars overall; a shorter star formation episode (top-left panel of Fig.~\ref{FigAvSFHSims}); and a faster loss of baryons (top panel of Fig.~\ref{FigFbEvol}).

\begin{figure*}
\begin{center}
\includegraphics[width=140mm]{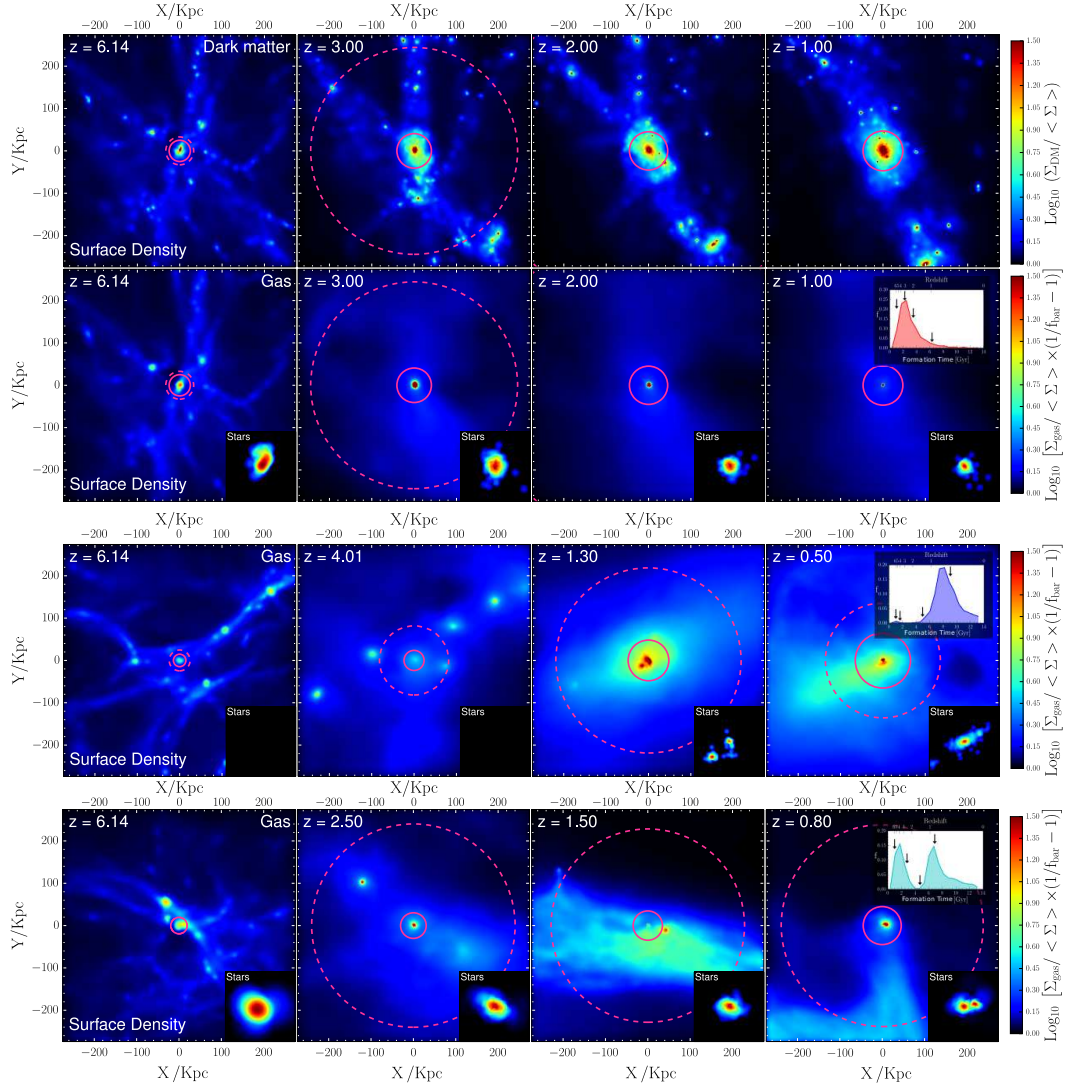}
\end{center}
 \caption{Evolution of the dark matter, gas, and stellar components of three dwarf galaxies with little intermediate start formation ($f_{\rm int}<0.2$). The top two rows correspond to a system with a single, old stellar population ($f_{\rm young}<0.1$, shown in red in Fig.~\ref{FigSimSFHT}). The two bottom rows show the gas and stars (dark matter is omitted, for clarity) of a system where recent star formation prevails ($f_{\rm old}<0.1$, shown in blue in Fig.~\ref{FigSimSFHT}), and one with two well-defined components of distinct age (cyan). In each panel the solid red circle indicated the virial radius of the central dwarf. The dotted circle indicates the smallest radius where the enclosed baryon fraction equals the universal average. Insets in the gas component panels show the stars formed within the virial radius of each dwarf. Extra insets in the rightmost panels show the star formation history of each dwarf.}
  \label{FigExamples}
\end{figure*}

The top two rows of Fig.~\ref{FigExamples} show the evolution of gas, dark matter, and stars in one such galaxy. The top row shows, in projection, the dark matter distribution within a box of $\sim 500$ kpc (comoving) on a side; the second row its corresponding gas component. Note how the gas traces the dark matter faithfully before reionization (leftmost panels) and how the gas is driven out of small halos soon after reionization is complete. The solid red circles in each panel indicate the virial radius of the central dwarf. 

The dotted circles in Fig.~\ref{FigExamples}, on the other hand, indicate the smallest volume centered on the dwarf where the enclosed baryon fraction equals the universal average. By $z=3$, reionization, aided by stellar feedback, have altered the baryon content of a region that extends as far as $\sim 7$ virial radii. This loss of baryons removes the fuel necessary to form stars at late times. Star formation, which had already begun by $z_{\rm reion}$, declines steadily after $z=3$ (see the inset in the rightmost panel of Fig.~\ref{FigExamples}) to negligible values at $z=0$.

A contrasting example is provided by dwarfs with a prevailing young stellar population (i.e., those with $f_{\rm int}<0.2$ and $f_{\rm old} <0.1$, shown in blue). These systems have completely different SFHs mostly because of their extremely low masses at $z_{\rm reion}$. As a result, gas in these systems has yet to start cooling and turning into stars when reionization occurs. Their virial temperatures are below $10^4$ K at $z_{\rm reion}$, and they are therefore unable to retain their reionization-heated baryons, which are quickly pushed out of the virial boundaries of the halos. 

Although these baryons are not completely lost, they do however become temporarily too tenuous to cool and assemble into galaxies. As the system grows more massive at later times, these baryons are eventually ``recaptured'', cool, and collapse to start forming a galaxy. Simulated dwarfs with a dominant young stellar population thus inhabit halos that collapsed late and that did not start forming stars in significant numbers until well after reionization. 

Interestingly, it is actually the lack of early star formation that minimizes the effects of feedback, thus helping to preserve the baryon content of these halos and allowing star formation to proceed at late times.  Cosmic reionization, in this case, acts mainly to delay star formation until the halo grows massive enough to reaccrete the heated baryons, allowing them to cool and collapse into a late-forming galaxy. The evolution of the gas component of one representative example of this group is shown in the third row of Fig.~\ref{FigExamples}. Note, in particular, how star formation is shifted to late times in this case, with the first stars appearing just before $z=2$ in this example.

It is important to note at this point that, although young stars dominate in these systems and star formation does not begin in earnest until $z\sim 2$ (see bottom panel of Fig.~\ref{FigFbEvol}), there is in every case a sprinkling of very old stars. Thus our simulations are not in conflict with the long-established observation that star formation started at very early times in {\it all} nearby dwarfs, regardless of their present-day mass or morphology \citep[see, e.g.,][]{Tolstoy2009}.

Finally, what about systems where both old and young stellar components co-exist at $z=0$ (shown in cyan)? These are actually produced by late mergers between prototypes of each of the two examples discussed in the preceding paragraph. In a typical case, a gas-depleted dwarf with a single, old stellar population merges at late times with a comparatively gas-rich, late-forming dwarf whose star formation has been delayed by reionization. The merger allows the remaining gas to cool efficiently, fostering a late episode of star formation.The bottom row of Fig.~\ref{FigExamples} shows the evolution of the gas component of a representative example of this group.

We conclude that the effects of cosmic reionization, acting on systems with virial temperatures of order $10^4$ $\mathrm{K}$ at $z_{\rm reion}$, provide a simple and compelling explanation of the major features of the star formation histories of nearby dwarfs.

\section{Discussion and Conclusions}
\label{SecConc}

The discussion above illustrates the complex interplay between mass accretion history and reionization, and how they determine, together with stellar feedback, the star formation history of a dwarf galaxy. This interplay leads to a qualitative but compelling scenario that may help to explain the bewildering diversity of star formation histories of nearby dwarfs, and to provide clues to the origin of the decline in star formation activity at intermediate times shown in Fig.~\ref{FigSFHObsDwarfs}.  The scenario relies on the dramatic effect that cosmic reionization has on the baryonic component of halos whose virial temperatures at $z_{\rm reion}$ are comparable to or lower than $\sim 10^4$ $\mathrm{K}$. This characteristic virial temperature defines a ``threshold'' mass at $z_{\rm reion}$ that largely determines the future SFH of the system.

Progenitors just above the threshold are able to initiate star formation before reionization, but their SFHs are sharply truncated by the loss of baryons induced by the combined effects of reionization and feedback. These systems typically evolve to form dwarfs characterized by a single population of ancient stars. 

Systems whose masses are well below the threshold at $z_{\rm reion}$ are unable to start forming stars before reionization, whose main effect is to delay the onset of their star forming activity until (and if) the system grows massive enough at later times. In either case star formation is disfavoured at intermediate times, leading to a gap in intermediate-age stars. This suggests that the intermediate-age hiatus in star formation activity observed in nearby dwarfs is a direct consequence of cosmic reionization.

Note that this explanation relies on a number of premises that are amenable to independent verification. One is that the nearby dwarfs in our sample (which span the range $ 10^5<M_{\rm gal}/M_\odot<10^9$) inhabit fairly low-mass halos, since their virial temperatures must have been at or below $\sim 10^4$ $\mathrm{K}$ at $z_{\rm reion}$. This constraint implies that their progenitors had, at $z_{\rm reion}$, a virial mass of order $\sim 2\times 10^9 \, M_\odot (1+z_{\rm reion})^{-3/2}$, or $\sim 1.2 \times 10^8 \, M_\odot$ for $z_{\rm reion}=6$. It is unlikely that these halos could have since grown by much more than a factor of $\sim 40$, implying virial masses at $z=0$ of roughly $5 \times 10^{9}\, M_\odot$ or lower. Note that these estimates would be shifted to lower values if, given the latest results of the WMAP and Planck missions, the reionization redshift is as high as $\sim 10$ \citep{Bennett2013,PlanckColl2013}.

Halo masses that low are at odds with abundance matching estimates, which indicate that nearby dwarfs in our sample should inhabit much more massive halos \citep[e.g.,][]{Guo2011,Moster2013}. Indeed, according to these models, a $10^8 \, M_\odot$ stellar mass dwarf should be surrounded by a $\sim 3\times 10^{10}\, M_\odot$ halo, roughly an order of magnitude more massive than our models would suggest if the intermediate star formation ``gap'' highlighted in Fig.~\ref{FigSFHObsDwarfs} is indeed a signature of cosmic reionization. This tension between models and observation is not new, and a number of recent papers have presented compelling evidence that the halo mass of dwarf galaxies is much lower than expected from abundance-matching \citep[see, e.g.,][]{Boylan-Kolchin2011,Parry2012,Ferrero2012}. There is still no consensus on how to resolve this discrepancy, with some authors arguing that it might require a major revision to the $\Lambda$CDM paradigm \citep[see, e.g.,][]{Bode2001,Vogelsberger2012} and 
others that it merely reflects the limitations of abundance-matching techniques when applied to the dwarf galaxy regime \citep{Sawala2014}. 

A second, independent line of enquiry that might help to validate the scenario we propose here concerns the origin of ``two-component'' dwarfs where prominent populations well separated in age coexist in the same galaxy. There are several known examples in the local Universe, including satellites of Andromeda and the Milky Way such as And II \citep{McConnachie2007}, Sculptor, Fornax and Sextans \citep{Tolstoy2004,Battaglia2006,Battaglia2011}, and nearby field dwarfs such as KKH37 and DDO6 \citep{Weisz2011}. The best studied examples indicate that the two components differ not only in age, but also in their metallicity, spatial distribution, and kinematics \citep{Walker2011}. These correlations place constraints on the formation paths allowed for these systems, and may thus be used to test the idea that cosmic reionization plays a substantial role in the origin of such distinct stellar components. We are currently investigating this issue, and plan to report on this topic in a future paper.

Our main conclusions may be summarized as follows.
\begin{itemize}
\item{The star formation histories of nearby field dwarfs show great diversity, but have also some well-defined ensemble characteristics. Taken as a group, these dwarfs have dominant old stellar populations; on average, dwarfs formed stars in the first 4 Gyr at a rate almost double their current past average, $\bar f=M_{\rm gal}/t_0$. They also exhibit a pronounced gap in their star forming activity at intermediate time; the average rate in the period $4<t/$Gyr$<8$ declines to less than one tenth of $\bar f$ before climbing back to roughly $0.6\, \bar f$ in the last $\sim 5$ Gyr.}
\item{Field dwarfs in the Local Group CLUES simulation have star formation histories as diverse as those of observed dwarfs. In particular, some simulated dwarfs show an intermediate-age ``gap'' reminiscent of that seen in nearby dwarfs. This gap is caused by cosmic reionization, which shuts off the gas supply of systems that have already started to form stars before $z_{\rm reion}$, and delays the onset of star formation in systems that assemble after $z_{\rm reion}$. }
\item{The simulated dwarfs where the intermediate gap is most evident are those whose progenitors, at $z_{\rm reion}$, had virial temperatures at or below $\sim 10^4$ $\mathrm{K}$. This corresponds, for $z_{\rm reion}=6$, to $M_{200}\simlt 1.2 \times 10^8\, M_\odot$. These progenitors grow to virial masses of order $\sim 5 \times 10^9\, M_\odot$ at $z=0$.  If the intermediate-age gap observed in the ensemble SFH of nearby dwarfs is indeed caused by reionization, this suggests that they inhabit halos that are nearly an order of magnitude less massive than than expected from abundance matching analysis. }
\end{itemize}

Our results thus provide indirect, but independent, support for the idea that dwarf galaxies inhabit halos of much lower mass than expected from simple models of galaxy formation. The sheer number of such halos predicted in the $\Lambda$CDM paradigm represent a challenge, especially because any successful model must explain why only a small fraction of them are populated today by luminous galaxies. However this challenge is eventually puzzled out, our results suggest that cosmic reionization is surely to emerge as an essential ingredient of the solution.

\section{Acknowledgements}
We thank Else Starkenburg, Evan Skillman, Daniel Weisz, Kim Venn and Igor Karachentsev for useful discussions. This work was supported in part by the National Science Foundation under Grant No. PHYS-1066293 and the hospitality of the Aspen Center for Physics.  The simulations were performed at Leibniz Rechenzentrum Munich (LRZ) and at Barcelona Supercomputing Center (BSC). Our collaboration has been supported by DFG grants GO 563/21-1 and GO 563/24-1 as well as by CONICET. ABL, JFN and MGA aknowledge support from ANPCyT grant PICT2012-1137. GY acknowledges financial support from MINECO under research grants AYA2012-31101, FPA2012-34694 and MultiDark CSD2009-00064. YH has been partially supported by the Israel Science Foundation (13/08).

\begin{table*}
\caption{Compilation of nearby {\it isolated} dwarf galaxies included in our analysis. We exclude systems brighter than $M_{\rm B}\approx -16$ and fainter than $M_{\rm B}\approx -7$, as well as satellites of the Milky Way, M31, M81, M82 and NGC 2403. See Sec.~\ref{SecObsDwarfs} for details on the selection. Galactic coordinates were taken from SIMBAD Astronomical Database. Heliocentric radial velocity, distances and magnitudes were taken from the public Catalog and Atlas of the LV galaxies (LVG) \citep[][and references therein]{Karachentsev2004, Karachentsev2013}. SFHs and stellar masses were taken from the references listed here. If the latter were not specified in the original reference, then we adopted the value listed in the LVG.}
\label{TabObsDwarfs}
{\scriptsize
\begin{tabular}{@{}lllllllllll}
\hline
Galaxy & $l$ & $b$ & $\mathrm{D_{\odot}}$ & $ \mathrm{V_{\odot}}$ & $\mathrm{M_{\rm gal}}$ & $\mathrm{M_B}$  &$f_{\rm old}$ & $f_{\rm int}$ & $f_{\rm young}$ & Reference \\
       & $(^o)$ & $(^o)$  & (Mpc) & $\mathrm{(km \ s^{-1})}$ &  ($\mathrm{10^7 \ M_{\odot}}$) & (mag)  & & & \\
\hline
LeoT 		& 214.85 & +43.66 & 0.42 & +39.0  & 0.01  & -6.73  & 0.547 & 0.347 & 0.105 & \citet{Weisz2012} \\
UA292 		& 148.28 & +83.72 & 3.61 & +308.0 & 0.13  & -11.79 & 0.551 & 0.000 & 0.449 & \citet{Weisz2011} \\
Cetus 		& 101.44 & -72.86 & 0.78 & -87.0  & 0.19  & -10.18 & 0.869 & 0.111 & 0.020 & \citet{Monelli2010a} \\
Tucana 		& 322.91 & -47.37 & 0.88 & +194.0 & 0.26  & -9.16  & 0.950 & 0.050 & 0.000 & \citet{Monelli2010b} \\
KKR25 		& 83.88  & +44.41 & 1.86 & -79.0  & 0.27  & -9.39  & 0.672 & 0.012 & 0.317 & \citet{Weisz2011} \\
KK230 		& 67.71  & +71.99 & 2.14 & +63.0  & 0.39  & -9.21  & 0.826 & 0.000 & 0.174 & \citet{Weisz2011} \\
Phoenix		& 272.16 & -68.95 & 0.44 & -52.0  & 0.44  & -9.56  & 0.562 & 0.318 & 0.120 & \citet{Hidalgo2009} \\
KKH86 		& 339.04 & +62.60 & 2.59 & +287.0 & 0.44  & -10.30 & 0.747 & 0.117 & 0.135 & \citet{Weisz2011} \\
KDG2 		& 119.78 & -80.94 & 3.40 & +224.0 & 0.49  & -11.39 & 0.030 & 0.000 & 0.970 & \citet{Weisz2011} \\
Antlia 		& 263.10 & +22.31 & 1.32 & +362.0 & 0.54  & -9.75  & 0.199 & 0.311 & 0.490 & \citet{Weisz2011} \\
E269-37 	& 303.59 & +17.03 & 3.48 & +744.0 & 0.77  & -11.39 & 0.959 & 0.000 & 0.041 & \citet{Weisz2011} \\
KKH98 		& 109.09 & -22.38 & 2.45 & -132.0 & 0.92  & -10.78 & 0.218 & 0.492 & 0.290 & \citet{Weisz2011} \\
KDG73 		& 136.88 & +44.23 & 3.70 & +116.0 & 0.94  & -10.76 & 0.391 & 0.000 & 0.609 & \citet{Weisz2011} \\
SagDIG 		& 21.06  & -16.28 & 1.04 & -79.0  & 0.96  & -11.49 & 0.050 & 0.471 & 0.479 & \citet{Held2007} \\
UGC4879		& 164.66 & +42.89 & 1.36 & -25.0  & 1.13  & -11.94 & 0.959 & 0.000 & 0.041 & \citet{Jacobs2011}  \\
GR8 		& 310.73 & +76.98 & 2.13 & +217.0 & 1.29  & -11.96 & 0.658 & 0.112 & 0.230 & \citet{Weisz2011} \\
DDO187 		& 25.57  & +70.46 & 2.20 & +160.0 & 1.37  & -12.44 & 0.547 & 0.024 & 0.429 & \citet{Weisz2011} \\
E294-010 	& 320.41 & -74.42 & 1.92 & +107.0 & 1.64  & -10.91 & 0.839 & 0.063 & 0.098 & \citet{Weisz2011} \\
Sc22 		& 52.74  & -83.34 & 4.21 & --     & 1.78  & -10.46 & 0.791 & 0.011 & 0.198 & \citet{Weisz2011} \\
DDO113 		& 161.10 & +78.06 & 2.86 &   --   & 1.81  & -11.51 & 0.674 & 0.000 & 0.326 & \citet{Weisz2011}  \\
E410-005 	& 357.85 & -80.71 & 1.92 & +36.0  & 2.00  & -11.58 & 0.667 & 0.183 & 0.151 & \citet{Weisz2011} \\
U8833 		& 67.71  & +73.96 & 3.08 & +221.0 & 2.18  & -12.20 & 0.783 & 0.011 & 0.206 & \citet{Weisz2011} \\
U4483 		& 144.97 & +34.38 & 3.21 & +156.0 & 2.23  & -12.73 & 0.000 & 0.938 & 0.062 & \citet{Weisz2011} \\
KDG52 		& 143.82 & +33.01 & 3.55 & +116.0 & 2.26  & -11.49 & 0.952 & 0.000 & 0.048 & \citet{Weisz2011} \\
KKH37 		& 133.98 & +26.54 & 3.39 & +11.0  & 2.32  & -11.58 & 0.544 & 0.036 & 0.419 & \citet{Weisz2011} \\
E321-014 	& 294.85 & +24.05 & 3.18 & +609.0 & 2.36  & -12.70 & 0.819 & 0.053 & 0.128 & \citet{Weisz2011} \\
DDO6 		& 119.39 & -83.88 & 3.34 & +292.0 & 2.50  & -12.39 & 0.647 & 0.000 & 0.353 & \citet{Weisz2011} \\
U8508 		& 111.14 & +61.31 & 2.69 & +56.0  & 3.26  & -13.09 & 0.674 & 0.000 & 0.326 & \citet{Weisz2011} \\
E540-032 	& 121.01 & -82.77 & 3.42 & +228.0 & 3.41  & -11.32 & 0.902 & 0.000 & 0.098 & \citet{Weisz2011} \\
DDO181 		& 89.73  & +73.12 & 3.01 & +214.0 & 3.67  & -13.20 & 0.794 & 0.000 & 0.206 & \citet{Weisz2011} \\
N3741 		& 157.57 & +66.45 & 3.03 & +229.0 & 3.77  & -13.13 & 0.761 & 0.000 & 0.239 & \citet{Weisz2011} \\
UA438 		& 11.87  & -70.86 & 2.18 & +62.0  & 3.83  & -12.86 & 0.647 & 0.298 & 0.055 & \citet{Weisz2011} \\
DDO183 		& 77.79  & +73.45 & 3.22 & +188.0 & 4.69  & -13.16 & 0.744 & 0.000 & 0.256 & \citet{Weisz2011} \\
DDO190 		& 82.01  & +64.48 & 2.74 & +150.0 & 6.24  & -14.14 & 0.398 & 0.112 & 0.490 & \citet{Weisz2011} \\
DDO53 		& 149.29 & +34.95 & 3.56 & +19.0  & 6.67  & -13.37 & 0.480 & 0.176 & 0.344 & \citet{Weisz2011} \\
DDO99 		& 166.20 & +72.44 & 2.64 & +251.0 & 7.84  & -13.52 & 0.768 & 0.184 & 0.048 & \citet{Weisz2011} \\
E325-11 	& 313.50 & +19.91 & 3.40 & +544.0 & 9.86  & -14.02 & 0.649 & 0.112 & 0.239 & \citet{Weisz2011} \\
IC1613 		& 129.74 & -60.58 & 0.73 & -232.0 & 10.00 & -14.54 & 0.361 & 0.229 & 0.410 & \citet{Skillman2003} \\
N4163 		& 163.21 & +77.70 & 2.94 & +162.0 & 12.51 & -13.80 & 0.462 & 0.483 & 0.055 & \citet{Weisz2011} \\
SexA 		& 246.15 & +39.88 & 1.32 & +324.0 & 13.83 & -13.93 & 0.366 & 0.195 & 0.439 & \citet{Weisz2011} \\
SexB 		& 233.20 & +43.78 & 1.36 & 300.0  & 16.52 & -13.95 & 0.656 & 0.000 & 0.344 & \citet{Weisz2011} \\
WLM 		& 75.85  & -73.63 & 0.97 & -122.0 & 18.60 & -14.06 & 0.596 & 0.199 & 0.206 & \citet{Dolphin2000} \\
LeoA 		& 196.90 & +52.42 & 0.81 & +24.0  & 20.00 & -11.70 & 0.176 & 0.303 & 0.521 & \citet{Cole2007} \\
N3109 		& 262.10 & +23.07 & 1.32 & +403.0 & 35.07 & -15.73 & 0.849 & 0.000 & 0.151 & \citet{Weisz2011} \\
DDO165 		& 120.75 & +49.36 & 4.57 & +31.0  & 43.06 & -15.09 & 0.635 & 0.012 & 0.353 & \citet{Weisz2011} \\
DDO125 		& 137.75 & +72.94 & 2.74 & +206.0 & 58.22 & -14.33 & 0.424 & 0.556 & 0.020 & \citet{Weisz2011} \\
\hline
\end{tabular}
}
\end{table*}

\bibliographystyle{mn2e}
\bibliography{my_biblio}

\begin{thebibliography}{}

\bibitem[\protect\citeauthoryear{{Babul} \& {Rees}}{{Babul} \&
  {Rees}}{1992}]{Babul1992}
{Babul} A.,  {Rees} M.~J.,  1992, \mnras, 255, 346

\bibitem[\protect\citeauthoryear{{Balogh}, {Navarro} \& {Morris}}{{Balogh}
  et~al.}{2000}]{Balogh2000}
{Balogh} M.~L.,  {Navarro} J.~F.,    {Morris} S.~L.,  2000, \apj, 540, 113

\bibitem[\protect\citeauthoryear{{Barkana} \& {Loeb}}{{Barkana} \&
  {Loeb}}{1999}]{Barkana1999}
{Barkana} R.,  {Loeb} A.,  1999, \apj, 523, 54

\bibitem[\protect\citeauthoryear{{Battaglia}, {Tolstoy}, {Helmi}, {Irwin},
  {Parisi}, {Hill} \& {Jablonka}}{{Battaglia} et~al.}{2011}]{Battaglia2011}
{Battaglia} G.,  {Tolstoy} E.,  {Helmi} A.,  {Irwin} M.,  {Parisi} P.,  {Hill}
  V.,    {Jablonka} P.,  2011, \mnras, 411, 1013

\bibitem[\protect\citeauthoryear{{Battaglia}, {Tolstoy}, {Helmi}, {Irwin},
  {Letarte}, {Jablonka}, {Hill}, {Venn}, {Shetrone}, {Arimoto}, {Primas},
  {Kaufer}, {Francois}, {Szeifert}, {Abel} \& {Sadakane}}{{Battaglia}
  et~al.}{2006}]{Battaglia2006}
{Battaglia} G.,  {Tolstoy} E.,  {Helmi} A.,  {Irwin} M.~J.,  {Letarte} B.,
  {Jablonka} P.,  {Hill} V.,  {Venn} K.~A.,  {Shetrone} M.~D.,  {Arimoto} N.,
  {Primas} F.,  {Kaufer} A.,  {Francois} P.,  {Szeifert} T.,  {Abel} T.,
  {Sadakane} K.,  2006, \aap, 459, 423

\bibitem[\protect\citeauthoryear{{Ben{\'{\i}}tez-Llambay}, {Navarro}, {Abadi},
  {Gottl{\"o}ber}, {Yepes}, {Hoffman} \& {Steinmetz}}{{Ben{\'{\i}}tez-Llambay}
  et~al.}{2013}]{Benitez-Llambay2013}
{Ben{\'{\i}}tez-Llambay} A.,  {Navarro} J.~F.,  {Abadi} M.~G.,  {Gottl{\"o}ber}
  S.,  {Yepes} G.,  {Hoffman} Y.,    {Steinmetz} M.,  2013, \apjl, 763, L41

\bibitem[\protect\citeauthoryear{{Bennett}, {Larson}, {Weiland}
  et~al.,}{{Bennett} et~al.}{2013}]{Bennett2013}
{Bennett} C.~L.,  {Larson} D.,  {Weiland} J.~L.,    et~al., 2013, \apjs, 208,
  20

\bibitem[\protect\citeauthoryear{{Benson}, {Lacey}, {Baugh}, {Cole} \&
  {Frenk}}{{Benson} et~al.}{2002}]{Benson2002}
{Benson} A.~J.,  {Lacey} C.~G.,  {Baugh} C.~M.,  {Cole} S.,    {Frenk} C.~S.,
  2002, \mnras, 333, 156

\bibitem[\protect\citeauthoryear{{Bode}, {Ostriker} \& {Turok}}{{Bode}
  et~al.}{2001}]{Bode2001}
{Bode} P.,  {Ostriker} J.~P.,    {Turok} N.,  2001, \apj, 556, 93

\bibitem[\protect\citeauthoryear{{Boylan-Kolchin}, {Bullock} \&
  {Kaplinghat}}{{Boylan-Kolchin} et~al.}{2011}]{Boylan-Kolchin2011}
{Boylan-Kolchin} M.,  {Bullock} J.~S.,    {Kaplinghat} M.,  2011, \mnras, 415,
  L40

\bibitem[\protect\citeauthoryear{{Brook}, {Di Cintio}, {Knebe},
  {Gottl{\"o}ber}, {Hoffman}, {Yepes} \& {Garrison-Kimmel}}{{Brook}
  et~al.}{2014}]{Brook2014}
{Brook} C.~B.,  {Di Cintio} A.,  {Knebe} A.,  {Gottl{\"o}ber} S.,  {Hoffman}
  Y.,  {Yepes} G.,    {Garrison-Kimmel} S.,  2014, \apjl, 784, L14

\bibitem[\protect\citeauthoryear{{Brown}, {Tumlinson}, {Geha}, {Kirby},
  {VandenBerg}, {Mu{\~n}oz}, {Kalirai}, {Simon}, {Avila}, {Guhathakurta},
  {Renzini} \& {Ferguson}}{{Brown} et~al.}{2012}]{Brown2012}
{Brown} T.~M.,  {Tumlinson} J.,  {Geha} M.,  {Kirby} E.~N.,  {VandenBerg}
  D.~A.,  {Mu{\~n}oz} R.~R.,  {Kalirai} J.~S.,  {Simon} J.~D.,  {Avila} R.~J.,
  {Guhathakurta} P.,  {Renzini} A.,    {Ferguson} H.~C.,  2012, \apjl, 753, L21

\bibitem[\protect\citeauthoryear{{Bullock}, {Kravtsov} \& {Weinberg}}{{Bullock}
  et~al.}{2000}]{Bullock2000}
{Bullock} J.~S.,  {Kravtsov} A.~V.,    {Weinberg} D.~H.,  2000, \apj, 539, 517

\bibitem[\protect\citeauthoryear{{Cole} \& {Lcid Team}}{{Cole} \& {Lcid
  Team}}{2007}]{Cole2007}
{Cole} A.~A.,  {Lcid Team} 2007, in {Vazdekis} A.,  {Peletier} R.,  eds, IAU
  Symposium Vol.~241 of IAU Symposium, {The ACS LCID Project: Quantifying the
  Delayed Star Formation in Leo A}.
pp 295--299

\bibitem[\protect\citeauthoryear{{Di Cintio}, {Knebe}, {Libeskind}, {Hoffman},
  {Yepes} \& {Gottl{\"o}ber}}{{Di Cintio} et~al.}{2012}]{DiCintio2012}
{Di Cintio} A.,  {Knebe} A.,  {Libeskind} N.~I.,  {Hoffman} Y.,  {Yepes} G.,
  {Gottl{\"o}ber} S.,  2012, \mnras, 423, 1883

\bibitem[\protect\citeauthoryear{{Dolphin}}{{Dolphin}}{2000}]{Dolphin2000}
{Dolphin} A.~E.,  2000, \apj, 531, 804

\bibitem[\protect\citeauthoryear{{Efstathiou}}{{Efstathiou}}{1992}]{Efstathiou1992}
{Efstathiou} G.,  1992, \mnras, 256, 43P

\bibitem[\protect\citeauthoryear{{Ferrero}, {Abadi}, {Navarro}, {Sales} \&
  {Gurovich}}{{Ferrero} et~al.}{2012}]{Ferrero2012}
{Ferrero} I.,  {Abadi} M.~G.,  {Navarro} J.~F.,  {Sales} L.~V.,    {Gurovich}
  S.,  2012, \mnras, 425, 2817

\bibitem[\protect\citeauthoryear{{Gill}, {Knebe}, {Gibson} \& {Dopita}}{{Gill}
  et~al.}{2004}]{Gill2004}
{Gill} S.~P.~D.,  {Knebe} A.,  {Gibson} B.~K.,    {Dopita} M.~A.,  2004,
  \mnras, 351, 410

\bibitem[\protect\citeauthoryear{{Gnedin}}{{Gnedin}}{2000}]{Gnedin2000}
{Gnedin} N.~Y.,  2000, \apj, 542, 535

\bibitem[\protect\citeauthoryear{{Gottloeber}, {Hoffman} \&
  {Yepes}}{{Gottloeber} et~al.}{2010}]{Gottloeber2010}
{Gottloeber} S.,  {Hoffman} Y.,    {Yepes} G.,  2010, ArXiv e-prints

\bibitem[\protect\citeauthoryear{{Grebel} \& {Gallagher} III}{{Grebel} \&
  {Gallagher}}{2004}]{Grebel2004}
{Grebel} E.~K.,  {Gallagher} III J.~S.,  2004, \apjl, 610, L89

\bibitem[\protect\citeauthoryear{{Guo}, {White}, {Boylan-Kolchin}, {De Lucia},
  {Kauffmann}, {Lemson}, {Li}, {Springel} \& {Weinmann}}{{Guo}
  et~al.}{2011}]{Guo2011}
{Guo} Q.,  {White} S.,  {Boylan-Kolchin} M.,  {De Lucia} G.,  {Kauffmann} G.,
  {Lemson} G.,  {Li} C.,  {Springel} V.,    {Weinmann} S.,  2011, \mnras, 413,
  101

\bibitem[\protect\citeauthoryear{{Guo}, {White}, {Li} \&
  {Boylan-Kolchin}}{{Guo} et~al.}{2010}]{Guo2010}
{Guo} Q.,  {White} S.,  {Li} C.,    {Boylan-Kolchin} M.,  2010, \mnras, 404,
  1111

\bibitem[\protect\citeauthoryear{{Haardt} \& {Madau}}{{Haardt} \&
  {Madau}}{1996}]{Haardt1996}
{Haardt} F.,  {Madau} P.,  1996, \apj, 461, 20

\bibitem[\protect\citeauthoryear{{Held}, {Momany}, {Rizzi}, {Saviane}, {Bedin},
  {Gullieuszik}, {Bertelli}, {Nasi}, {Clemens}, {Rich} \& {Kuijken}}{{Held}
  et~al.}{2007}]{Held2007}
{Held} E.~V.,  {Momany} Y.,  {Rizzi} L.,  {Saviane} I.,  {Bedin} L.~R.,
  {Gullieuszik} M.,  {Bertelli} G.,  {Nasi} E.,  {Clemens} M.,  {Rich} M.~R.,
   {Kuijken} K.,  2007, in {Vazdekis} A.,  {Peletier} R.,  eds, IAU Symposium
  Vol.~241 of IAU Symposium, {The star formation history of the dwarf irregular
  galaxy SagDIG}.
pp 339--340

\bibitem[\protect\citeauthoryear{{Hidalgo}, {Aparicio},
  {Mart{\'{\i}}nez-Delgado} \& {Gallart}}{{Hidalgo} et~al.}{2009}]{Hidalgo2009}
{Hidalgo} S.~L.,  {Aparicio} A.,  {Mart{\'{\i}}nez-Delgado} D.,    {Gallart}
  C.,  2009, \apj, 705, 704

\bibitem[\protect\citeauthoryear{{Hoeft}, {Yepes}, {Gottl{\"o}ber} \&
  {Springel}}{{Hoeft} et~al.}{2006}]{Hoeft2006}
{Hoeft} M.,  {Yepes} G.,  {Gottl{\"o}ber} S.,    {Springel} V.,  2006, \mnras,
  371, 401

\bibitem[\protect\citeauthoryear{{Jacobs}, {Tully}, {Rizzi}, {Karachentsev},
  {Chiboucas} \& {Held}}{{Jacobs} et~al.}{2011}]{Jacobs2011}
{Jacobs} B.~A.,  {Tully} R.~B.,  {Rizzi} L.,  {Karachentsev} I.~D.,
  {Chiboucas} K.,    {Held} E.~V.,  2011, \aj, 141, 106

\bibitem[\protect\citeauthoryear{{Karachentsev}, {Karachentseva}, {Huchtmeier}
  \& {Makarov}}{{Karachentsev} et~al.}{2004}]{Karachentsev2004}
{Karachentsev} I.~D.,  {Karachentseva} V.~E.,  {Huchtmeier} W.~K.,    {Makarov}
  D.~I.,  2004, \aj, 127, 2031

\bibitem[\protect\citeauthoryear{{Karachentsev}, {Makarov} \&
  {Kaisina}}{{Karachentsev} et~al.}{2013}]{Karachentsev2013}
{Karachentsev} I.~D.,  {Makarov} D.~I.,    {Kaisina} E.~I.,  2013, \aj, 145,
  101

\bibitem[\protect\citeauthoryear{{Katz}, {Weinberg} \& {Hernquist}}{{Katz}
  et~al.}{1996}]{Katz1996}
{Katz} N.,  {Weinberg} D.~H.,    {Hernquist} L.,  1996, \apjs, 105, 19

\bibitem[\protect\citeauthoryear{{Klypin}, {Kravtsov}, {Valenzuela} \&
  {Prada}}{{Klypin} et~al.}{1999}]{Klypin1999}
{Klypin} A.,  {Kravtsov} A.~V.,  {Valenzuela} O.,    {Prada} F.,  1999, \apj,
  522, 82

\bibitem[\protect\citeauthoryear{{Knebe}, {Libeskind}, {Knollmann},
  {Martinez-Vaquero}, {Yepes}, {Gottl{\"o}ber} \& {Hoffman}}{{Knebe}
  et~al.}{2011}]{Knebe2011}
{Knebe} A.,  {Libeskind} N.~I.,  {Knollmann} S.~R.,  {Martinez-Vaquero} L.~A.,
  {Yepes} G.,  {Gottl{\"o}ber} S.,    {Hoffman} Y.,  2011, \mnras, 412, 529

\bibitem[\protect\citeauthoryear{{Libeskind} et~al.,}{{Libeskind}
  et~al.}{2010}]{Libeskind2010}
{Libeskind} N.~I.,  et~al., 2010, \mnras, 401, 1889

\bibitem[\protect\citeauthoryear{{Ludlow}, {Navarro}, {Springel},
  {Vogelsberger}, {Wang}, {White}, {Jenkins} \& {Frenk}}{{Ludlow}
  et~al.}{2010}]{Ludlow2010}
{Ludlow} A.~D.,  {Navarro} J.~F.,  {Springel} V.,  {Vogelsberger} M.,  {Wang}
  J.,  {White} S.~D.~M.,  {Jenkins} A.,    {Frenk} C.~S.,  2010, \mnras, 406,
  137

\bibitem[\protect\citeauthoryear{{McConnachie}, {Pe{\~n}arrubia} \&
  {Navarro}}{{McConnachie} et~al.}{2007}]{McConnachie2007}
{McConnachie} A.~W.,  {Pe{\~n}arrubia} J.,    {Navarro} J.~F.,  2007, \mnras,
  380, L75

\bibitem[\protect\citeauthoryear{{Monelli}, {Gallart}, {Hidalgo}, {Aparicio},
  {Skillman}, {Cole}, {Weisz}, {Mayer}, {Bernard}, {Cassisi}, {Dolphin},
  {Drozdovsky} \& {Stetson}}{{Monelli} et~al.}{2010}]{Monelli2010b}
{Monelli} M.,  {Gallart} C.,  {Hidalgo} S.~L.,  {Aparicio} A.,  {Skillman}
  E.~D.,  {Cole} A.~A.,  {Weisz} D.~R.,  {Mayer} L.,  {Bernard} E.~J.,
  {Cassisi} S.,  {Dolphin} A.~E.,  {Drozdovsky} I.,    {Stetson} P.~B.,  2010,
  \apj, 722, 1864

\bibitem[\protect\citeauthoryear{{Monelli}, {Hidalgo}, {Stetson}, {Aparicio},
  {Gallart}, {Dolphin}, {Cole}, {Weisz}, {Skillman}, {Bernard}, {Mayer},
  {Navarro}, {Cassisi}, {Drozdovsky} \& {Tolstoy}}{{Monelli}
  et~al.}{2010}]{Monelli2010a}
{Monelli} M.,  {Hidalgo} S.~L.,  {Stetson} P.~B.,  {Aparicio} A.,  {Gallart}
  C.,  {Dolphin} A.~E.,  {Cole} A.~A.,  {Weisz} D.~R.,  {Skillman} E.~D.,
  {Bernard} E.~J.,  {Mayer} L.,  {Navarro} J.~F.,  {Cassisi} S.,  {Drozdovsky}
  I.,    {Tolstoy} E.,  2010, \apj, 720, 1225

\bibitem[\protect\citeauthoryear{{Moore}, {Ghigna}, {Governato}, {Lake},
  {Quinn}, {Stadel} \& {Tozzi}}{{Moore} et~al.}{1999}]{Moore1999}
{Moore} B.,  {Ghigna} S.,  {Governato} F.,  {Lake} G.,  {Quinn} T.,  {Stadel}
  J.,    {Tozzi} P.,  1999, \apjl, 524, L19

\bibitem[\protect\citeauthoryear{{Moster}, {Naab} \& {White}}{{Moster}
  et~al.}{2013}]{Moster2013}
{Moster} B.~P.,  {Naab} T.,    {White} S.~D.~M.,  2013, \mnras, 428, 3121

\bibitem[\protect\citeauthoryear{{Navarro} \& {Steinmetz}}{{Navarro} \&
  {Steinmetz}}{1997}]{Navarro1997}
{Navarro} J.~F.,  {Steinmetz} M.,  1997, \apj, 478, 13

\bibitem[\protect\citeauthoryear{{Okamoto} \& {Frenk}}{{Okamoto} \&
  {Frenk}}{2009}]{Okamoto2009}
{Okamoto} T.,  {Frenk} C.~S.,  2009, \mnras, 399, L174

\bibitem[\protect\citeauthoryear{{Okamoto}, {Gao} \& {Theuns}}{{Okamoto}
  et~al.}{2008}]{Okamoto2008}
{Okamoto} T.,  {Gao} L.,    {Theuns} T.,  2008, \mnras, 390, 920

\bibitem[\protect\citeauthoryear{{Orban}, {Gnedin}, {Weisz}, {Skillman},
  {Dolphin} \& {Holtzman}}{{Orban} et~al.}{2008}]{Orban2008}
{Orban} C.,  {Gnedin} O.~Y.,  {Weisz} D.~R.,  {Skillman} E.~D.,  {Dolphin}
  A.~E.,    {Holtzman} J.~A.,  2008, \apj, 686, 1030

\bibitem[\protect\citeauthoryear{{Parry}, {Eke}, {Frenk} \& {Okamoto}}{{Parry}
  et~al.}{2012}]{Parry2012}
{Parry} O.~H.,  {Eke} V.~R.,  {Frenk} C.~S.,    {Okamoto} T.,  2012, \mnras,
  419, 3304

\bibitem[\protect\citeauthoryear{{Planck Collaboration}, {Ade}, {Aghanim},
  {Armitage-Caplan}, {Arnaud}, {Ashdown}, {Atrio-Barandela}, {Aumont},
  {Baccigalupi}, {Banday} \& et al.}{{Planck Collaboration}
  et~al.}{2013}]{PlanckColl2013}
{Planck Collaboration} {Ade} P.~A.~R.,  {Aghanim} N.,  {Armitage-Caplan} C.,
  {Arnaud} M.,  {Ashdown} M.,  {Atrio-Barandela} F.,  {Aumont} J.,
  {Baccigalupi} C.,  {Banday} A.~J.,    et al. 2013, ArXiv e-prints: 1303.5076

\bibitem[\protect\citeauthoryear{{Ricotti}}{{Ricotti}}{2009}]{Ricotti2009}
{Ricotti} M.,  2009, \mnras, 392, L45

\bibitem[\protect\citeauthoryear{{Ricotti} \& {Gnedin}}{{Ricotti} \&
  {Gnedin}}{2005}]{Ricotti2005}
{Ricotti} M.,  {Gnedin} N.~Y.,  2005, \apj, 629, 259

\bibitem[\protect\citeauthoryear{{Sales}, {Helmi}, {Abadi}, {Brook},
  {G{\'o}mez}, {Ro{\v s}kar}, {Debattista}, {House}, {Steinmetz} \&
  {Villalobos}}{{Sales} et~al.}{2009}]{Sales2009}
{Sales} L.~V.,  {Helmi} A.,  {Abadi} M.~G.,  {Brook} C.~B.,  {G{\'o}mez} F.~A.,
   {Ro{\v s}kar} R.,  {Debattista} V.~P.,  {House} E.,  {Steinmetz} M.,
  {Villalobos} {\'A}.,  2009, \mnras, 400, L61

\bibitem[\protect\citeauthoryear{{Sawala}, {Frenk}, {Fattahi}, {Navarro},
  {Bower}, {Crain}, {Dalla Vecchia}, {Furlong}, {Jenkins}, {McCarthy}, {Qu},
  {Schaller}, {Schaye} \& {Theuns}}{{Sawala} et~al.}{2014}]{Sawala2014}
{Sawala} T.,  {Frenk} C.~S.,  {Fattahi} A.,  {Navarro} J.~F.,  {Bower} R.~G.,
  {Crain} R.~A.,  {Dalla Vecchia} C.,  {Furlong} M.,  {Jenkins} A.,  {McCarthy}
  I.~G.,  {Qu} Y.,  {Schaller} M.,  {Schaye} J.,    {Theuns} T.,  2014, ArXiv
  e-prints: 1404.3724

\bibitem[\protect\citeauthoryear{{Sawala}, {Guo}, {Scannapieco}, {Jenkins} \&
  {White}}{{Sawala} et~al.}{2011}]{Sawala2011}
{Sawala} T.,  {Guo} Q.,  {Scannapieco} C.,  {Jenkins} A.,    {White} S.,  2011,
  \mnras, 413, 659

\bibitem[\protect\citeauthoryear{{Sawala}, {Scannapieco}, {Maio} \&
  {White}}{{Sawala} et~al.}{2010}]{Sawala2010}
{Sawala} T.,  {Scannapieco} C.,  {Maio} U.,    {White} S.,  2010, \mnras, 402,
  1599

\bibitem[\protect\citeauthoryear{{Sawala}, {Scannapieco} \& {White}}{{Sawala}
  et~al.}{2012}]{Sawala2012}
{Sawala} T.,  {Scannapieco} C.,    {White} S.,  2012, \mnras, 420, 1714

\bibitem[\protect\citeauthoryear{{Shapiro}, {Giroux} \& {Babul}}{{Shapiro}
  et~al.}{1994}]{Shapiro1994}
{Shapiro} P.~R.,  {Giroux} M.~L.,    {Babul} A.,  1994, \apj, 427, 25

\bibitem[\protect\citeauthoryear{{Shen}, {Madau}, {Conroy}, {Governato} \&
  {Mayer}}{{Shen} et~al.}{2013}]{Shen2013}
{Shen} S.,  {Madau} P.,  {Conroy} C.,  {Governato} F.,    {Mayer} L.,  2013,
  ArXiv e-prints: 1308.4131

\bibitem[\protect\citeauthoryear{{Simpson}, {Bryan}, {Johnston}, {Smith}, {Mac
  Low}, {Sharma} \& {Tumlinson}}{{Simpson} et~al.}{2013}]{Simpson2013}
{Simpson} C.~M.,  {Bryan} G.~L.,  {Johnston} K.~V.,  {Smith} B.~D.,  {Mac Low}
  M.-M.,  {Sharma} S.,    {Tumlinson} J.,  2013, \mnras, 432, 1989

\bibitem[\protect\citeauthoryear{{Skillman}, {Hidalgo}, {Weisz}, {Monelli},
  {Gallart}, {Aparicio}, {Bernard}, {Boylan-Kolchin}, {Cassisi}, {Cole},
  {Dolphin}, {Ferguson}, {Mayer}, {Navarro}, {Stetson} \& {Tolstoy}}{{Skillman}
  et~al.}{2014}]{Skillman2014}
{Skillman} E.~D.,  {Hidalgo} S.~L.,  {Weisz} D.~R.,  {Monelli} M.,  {Gallart}
  C.,  {Aparicio} A.,  {Bernard} E.~J.,  {Boylan-Kolchin} M.,  {Cassisi} S.,
  {Cole} A.~A.,  {Dolphin} A.~E.,  {Ferguson} H.~C.,  {Mayer} L.,  {Navarro}
  J.~F.,  {Stetson} P.~B.,    {Tolstoy} E.,  2014, \apj, 786, 44

\bibitem[\protect\citeauthoryear{{Skillman}, {Tolstoy}, {Cole}, {Dolphin},
  {Saha}, {Gallagher}, {Dohm-Palmer} \& {Mateo}}{{Skillman}
  et~al.}{2003}]{Skillman2003}
{Skillman} E.~D.,  {Tolstoy} E.,  {Cole} A.~A.,  {Dolphin} A.~E.,  {Saha} A.,
  {Gallagher} J.~S.,  {Dohm-Palmer} R.~C.,    {Mateo} M.,  2003, \apj, 596, 253

\bibitem[\protect\citeauthoryear{{Somerville}}{{Somerville}}{2002}]{Somerville2002}
{Somerville} R.~S.,  2002, \apjl, 572, L23

\bibitem[\protect\citeauthoryear{{Spergel}, {Bean}, {Dor{\'e}}
  et~al.,}{{Spergel} et~al.}{2007}]{Spergel2007}
{Spergel} D.~N.,  {Bean} R.,  {Dor{\'e}} O.,    et~al., 2007, \apjs, 170, 377

\bibitem[\protect\citeauthoryear{{Springel}}{{Springel}}{2005}]{Springel2005}
{Springel} V.,  2005, \mnras, 364, 1105

\bibitem[\protect\citeauthoryear{{Springel} \& {Hernquist}}{{Springel} \&
  {Hernquist}}{2003}]{Springel2003}
{Springel} V.,  {Hernquist} L.,  2003, \mnras, 339, 289

\bibitem[\protect\citeauthoryear{{Springel}, {White}, {Tormen} \&
  {Kauffmann}}{{Springel} et~al.}{2001}]{Springel2001}
{Springel} V.,  {White} S.~D.~M.,  {Tormen} G.,    {Kauffmann} G.,  2001,
  \mnras, 328, 726

\bibitem[\protect\citeauthoryear{{Starkenburg}, {Helmi}, {De Lucia}, {Li},
  {Navarro}, {Font}, {Frenk}, {Springel}, {Vera-Ciro} \& {White}}{{Starkenburg}
  et~al.}{2013}]{Starkenburg2013}
{Starkenburg} E.,  {Helmi} A.,  {De Lucia} G.,  {Li} Y.-S.,  {Navarro} J.~F.,
  {Font} A.~S.,  {Frenk} C.~S.,  {Springel} V.,  {Vera-Ciro} C.~A.,    {White}
  S.~D.~M.,  2013, \mnras, 429, 725

\bibitem[\protect\citeauthoryear{{Susa} \& {Umemura}}{{Susa} \&
  {Umemura}}{2004}]{Susa2004}
{Susa} H.,  {Umemura} M.,  2004, \apj, 600, 1

\bibitem[\protect\citeauthoryear{{Thoul} \& {Weinberg}}{{Thoul} \&
  {Weinberg}}{1996}]{Thoul1996}
{Thoul} A.~A.,  {Weinberg} D.~H.,  1996, \apj, 465, 608

\bibitem[\protect\citeauthoryear{{Tolstoy}, {Hill} \& {Tosi}}{{Tolstoy}
  et~al.}{2009}]{Tolstoy2009}
{Tolstoy} E.,  {Hill} V.,    {Tosi} M.,  2009, \araa, 47, 371

\bibitem[\protect\citeauthoryear{{Tolstoy}, {Irwin}, {Helmi}, {Battaglia},
  {Jablonka}, {Hill}, {Venn}, {Shetrone}, {Letarte}, {Cole}, {Primas},
  {Francois}, {Arimoto}, {Sadakane}, {Kaufer}, {Szeifert} \& {Abel}}{{Tolstoy}
  et~al.}{2004}]{Tolstoy2004}
{Tolstoy} E.,  {Irwin} M.~J.,  {Helmi} A.,  {Battaglia} G.,  {Jablonka} P.,
  {Hill} V.,  {Venn} K.~A.,  {Shetrone} M.~D.,  {Letarte} B.,  {Cole} A.~A.,
  {Primas} F.,  {Francois} P.,  {Arimoto} N.,  {Sadakane} K.,  {Kaufer} A.,
  {Szeifert} T.,    {Abel} T.,  2004, \apjl, 617, L119

\bibitem[\protect\citeauthoryear{{Vogelsberger}, {Zavala} \&
  {Loeb}}{{Vogelsberger} et~al.}{2012}]{Vogelsberger2012}
{Vogelsberger} M.,  {Zavala} J.,    {Loeb} A.,  2012, \mnras, 423, 3740

\bibitem[\protect\citeauthoryear{{Walker} \& {Pe{\~n}arrubia}}{{Walker} \&
  {Pe{\~n}arrubia}}{2011}]{Walker2011}
{Walker} M.~G.,  {Pe{\~n}arrubia} J.,  2011, \apj, 742, 20

\bibitem[\protect\citeauthoryear{{Weisz} et~al.,}{{Weisz}
  et~al.}{2011}]{Weisz2011}
{Weisz} D.~R.,  et~al., 2011, \apj, 739, 5

\bibitem[\protect\citeauthoryear{{Weisz}, {Zucker}, {Dolphin}, {Martin}, {de
  Jong}, {Holtzman}, {Dalcanton}, {Gilbert}, {Williams}, {Bell}, {Belokurov} \&
  {Wyn Evans}}{{Weisz} et~al.}{2012}]{Weisz2012}
{Weisz} D.~R.,  {Zucker} D.~B.,  {Dolphin} A.~E.,  {Martin} N.~F.,  {de Jong}
  J.~T.~A.,  {Holtzman} J.~A.,  {Dalcanton} J.~J.,  {Gilbert} K.~M.,
  {Williams} B.~F.,  {Bell} E.~F.,  {Belokurov} V.,    {Wyn Evans} N.,  2012,
  \apj, 748, 88

\bibitem[\protect\citeauthoryear{{White} \& {Rees}}{{White} \&
  {Rees}}{1978}]{White1978}
{White} S.~D.~M.,  {Rees} M.~J.,  1978, \mnras, 183, 341

\bibitem[\protect\citeauthoryear{{Yepes}, {Gottl{\"o}ber} \& {Hoffman}}{{Yepes}
  et~al.}{2014}]{Yepes2014}
{Yepes} G.,  {Gottl{\"o}ber} S.,    {Hoffman} Y.,  2014, \nar, 58, 1

\end{thebibliography}
\end{document}